\documentclass[twocolumn]{aastex62}

\usepackage{lineno}
\newcommand{\teff}{\mbox{T$_{\rm eff}$}}
\newcommand{\logg}{\mbox{log~{\it g}}}
\newcommand{\vmicro}{\mbox{$\xi_{\rm t}$}}
\newcommand{\kmsec}{\mbox{km~s$^{\rm -1}$}}

\newcommand{\x}{\mbox{$\Delta_{\tiny{\mathrm{F275W,F814W}}}$}}
\newcommand{\y}{\mbox{$\Delta_{\tiny{C~\mathrm{ F275W,F336W,F438W}}}$}}

\received{19-Jul-2023}
\accepted{21-Sep-2023}
\submitjournal{ApJ}
\shorttitle{Chemical abundances in NGC\,104} 
\shortauthors{A.\,F. Marino, et al.} 

\begin{document}

\title{The metallicity variations along the chromosome maps: The Globular Cluster 47~Tucanae\footnote{Based on observations collected at the European Organisation for Astronomical Research in the Southern Hemisphere under ESO programme 105.20NB, and on observations with the NASA/ESA {\it Hubble Space Telescope}, obtained at the Space Telescope Science Institute, which is operated by AURA, Inc., under NASA contract NAS 5-26555. }}

\author{A.\ F.\,Marino}
\affiliation{Istituto Nazionale di Astrofisica - Osservatorio Astronomico di Padova, Vicolo dell'Osservatorio 5, Padova, IT-35122} 
\affiliation{Istituto Nazionale di Astrofisica - Osservatorio Astrofisico di Arcetri, Largo Enrico Fermi, 5, Firenze, IT-50125}
\author{A.\ P.\,Milone}
\affiliation{Dipartimento di Fisica e Astronomia ``Galileo Galilei'' - Univ. di Padova, Vicolo dell'Osservatorio 3, Padova, IT-35122}
\affiliation{Istituto Nazionale di Astrofisica - Osservatorio Astronomico di Padova, Vicolo dell'Osservatorio 5, Padova, IT-35122} 
\author{E.\ Dondoglio}
\affiliation{Dipartimento di Fisica e Astronomia ``Galileo Galilei'' - Univ. di Padova, Vicolo dell'Osservatorio 3, Padova, IT-35122}
\author{A.\ Renzini}
\affiliation{Istituto Nazionale di Astrofisica - Osservatorio Astronomico di Padova, Vicolo dell'Osservatorio 5, Padova, IT-35122} 
\author{G.\ Cordoni}
\affiliation{Istituto Nazionale di Astrofisica - Osservatorio Astronomico di Padova, Vicolo dell'Osservatorio 5, Padova, IT-35122}
\author{H.\ Jerjen}
\affiliation{Research School of Astronomy \& Astrophysics, Australian National University, Canberra, ACT 2611, Australia}  
\author{A.\ Karakas}
\affiliation{School of Physics and Astronomy, Monash University, VIC 3800, Australia} 
\affiliation{Centre of Excellence for Astrophysics in Three Dimensions (ASTRO-3D), Melbourne, Victoria, Australia}
\author{E.\ P.\,Lagioia}
\affiliation{South-Western Institute for Astronomy Research, Yunnan University, Kunming, 650500 P.R. China} 
\affiliation{Dipartimento di Fisica e Astronomia ``Galileo Galilei'' - Univ. di Padova, Vicolo dell'Osservatorio 3, Padova, IT-35122} 
\author{M.\ V.\,Legnardi}
\affiliation{Dipartimento di Fisica e Astronomia ``Galileo Galilei'' - Univ. di Padova, Vicolo dell'Osservatorio 3, Padova, IT-35122}
\author{M.\ McKenzie}
\affiliation{Research School of Astronomy \& Astrophysics, Australian National University, Canberra, ACT 2611, Australia} 
\author{A.\ Mohandasan}
\affiliation{Dipartimento di Fisica e Astronomia ``Galileo Galilei'' - Univ. di Padova, Vicolo dell'Osservatorio 3, Padova, IT-35122}
\author{M.\ Tailo}
\affiliation{Dipartimento di Fisica e Astronomia Augusto Righi, Università degli Studi di Bologna, Via Gobetti 93/2, I-40129, Bologna, Italy}
\author{D.\ Yong}
\affiliation{Research School of Astronomy \& Astrophysics, Australian National University, Canberra, ACT 2611, Australia} 
\author{T.\ Ziliotto}
\affiliation{Dipartimento di Fisica e Astronomia ``Galileo Galilei'' - Univ. di Padova, Vicolo dell'Osservatorio 3, Padova, IT-35122}

\correspondingauthor{A.\ F.\,Marino}
\email{anna.marino@inaf.it}

\begin{abstract}
The "chromosome maps" (ChMs) of globular clusters (GCs) have revealed
that these ancient structures are not homogeneous in metallicity in
various ways, and in different natures. The Type~II GCs, generally
display larger variations, sometimes coupled with slow neutron capture
($s$) element enrichment on the ChMs redder sequences, which has been
interpreted as due to multiple generations of stars. On the other
hand, most GCs have inhomogeneous first populations (1P) in the form
of large ranges in the \x\ values, pointing towards a not fully mixed
pristine molecular cloud. We analyse the chemical composition the GC
47~Tucanae, which shows both inhomogeneous 1P stars and, although not
formally a Type~II GC, hosts a small number of stars distributed on a
red side of the main stream of ChM stars. Our results suggest that 1P
stars are not homogeneous in the overall metallicity, with variations
of the order of $\sim$0.10~dex in all the chemical species. The {\it
  anomalous} stars distributed on a redder sequence of the ChM, are
further enriched in metals, but without any evidence for a significant
enrichment in the $s$ elements. Our three second population stars
located on the {\it normal} component of the map, have metallicities
similar to those of the metal-richer 1P group, suggesting that this
population formed from these stars. Although three stars is a
too-small sample to draw strong conclusions, the low spread in metals
of these objects might point towards a formation in a fully mixed
medium, possibly after a cooling flow phase. 
\end{abstract}


\section{Introduction}\label{sec:intro}

The existence of multiple stellar populations in virtually all
globular clusters (GCs) of the Milky Way is a well-established fact
\citep[see][for reviews]{Gratton12, Bastian15, MilMar22}. Yet, the
mechanisms responsible for this phenomenon continue to be debated
\citep[e.g.][]{Renzini2015}.   

The most effective tool to isolate the different populations of stars
hosted in a GC is offered by a combination of four bands from the
Hubble Space Telescope ($HST$), namely F275W,
F336W, F438W and F814W, allowing the construction of multiple
pseudo-color plots now known as ``Chromosome Maps''  
\citep[ChMs,][]{Mil15}.
In these diagrams the position of stars is especially sensitive to the
abundance of C, N and O elements via the molecules that they form as
well as to the helium and iron abundance \citep[e.g.][]{Mar08, Mil17,
  Mar19a}. 
The ChMs have revolutionized the traditional picture of GCs, and have
revealed unexpected features of the multiple population pattern.  

On the basis of the ChM shape, GCs have been subdivided into two main
groups, namely Type~I and Type~II. The majority ($\sim$83\%) of Milky
Way GCs belongs to the Type~I class, with the ChM displaying a first
population (1P) and different numbers of second population (2P)
groups, but following a single ``sequence'' along the map. ChMs of
Type~II clusters are more complex. In addition to the common 1P and 2P
stars, a second sequence (hereafter {\it anomalous}) appears on 
the red side of their maps \citep{Mil17}.  
A large number of  spectroscopic studies have demonstrated that stars
on the {\it anomalous} sequence are metal richer than the {\it normal}
ones and, in most cases, they also are enriched in the elements
produced by the slow neutron capture reactions 
($s$ elements) \citep[e.g.][]{Yong08, Mar09, Mar15, Mar21, Johnson17, McKenzie22}.

Perhaps the most shocking discovery from the ChMs was the apparent
chemical inhomogeneity of even the 1P, with elemental ratios similar
to that of halo field stars.  
Indeed, in most GCs, the 1P stars themselves do not look as a
chemically homogeneous group as they show a large spread in the
$m_{\rm F275W}-m_{\rm F814W}$ color, the $x$ axis of the ChMs.  
This was first noticed by \citet{Mil15} who hinted at a
variation of helium among 1P stars to account for the spread. 
However, none of the several plausible physical mechanisms considered
by \citet{Mil18} was able to account for an helium enrichment without
a concomitant conversion of C and O into N. 
Thus, either a totally new, unknown process was at work producing a
spread in helium among 1P stars, or helium was not the culprit and
some other element was driving it.  

To shed light on the apparent chemical inhomogeneity of 1P stars, once
associated to one single stellar population (the primordial one),
spectroscopic tagging in GCs with more pronounced spreads along the 1P
on the ChM has promptly followed the photometric findings. 
The first spectroscopic evidence of metallicity variations among 1P
stars has been shown in the GC NGC\,3201 ([Fe/H]=$-$1.59~dex), where a
small spread of the order of $\sim$0.10~dex has been detected, at
constant [O/Fe] and [Na/Fe] \citep{Mar19b}. 
These findings suggest indeed that the overall metallicity, rather than helium, is the main factor for the 1P spread along the ChM.
Later on, photometric support for internal variations in metallicity
among the 1P stars has been reported for several analysed GCs
\citep[see][]{Legnardi22}. 

The possibility that 1P stars themselves have an internal spread in
the overall metallicity suggests that either the molecular cloud out
of which 1P stars formed was not chemically homogeneous, i.e., was not
fully mixed following its original chemical enrichment,  
or while 1P stars were forming some Type~Ia supernova from previous
generations may have polluted the intra-cluster medium. 
These options can provide us with new hints to understand the process
of GC formation, its time-scale, and help to identify the class of
stars responsible for the further enrichment of the interstellar
medium (ISM) to form the 2P stars in GCs.  
In this context, it is noteworthy that hydrodynamical simulations have
predicted that metal abundance spreads within GC forming giant
molecular clouds can influence the iron abundances of future cluster
members \citep{McKenzie21}.  

It is clear now that the chemical enrichment in metallicity, though
small, might be a quite common property of Milky Way GCs. The
phenomenon seems to be linked to different mechanisms, depending on
whether the GC is Type~I, and metallicity variations could account for
the 1P spread in the $m_{\rm F275W}-m_{\rm F814W}$ color along the
ChMs, or Type~II where the enrichment of the redder {\it anomalous}
component could be larger, sometimes associated with further
enhancements in the [$s$/Fe] and C+N+O abundances
\citep[e.g.][]{Mar11a, Yong15}.  
In this context, we note that, when spectra of excellent quality are
used, very small variations in metals, of the order of a few
hundredths of dex are found in both globular and open clusters
\citep[e.g.][]{Yong13, Liu16, Monty23, Lardo23}. 

In this work we analyse the chemical pattern of 1P stars in the
relatively metal rich GC NGC\,104 (47~Tucanae, [Fe/H]=$-$0.72~dex)
\citep{Harris10}. This bright cluster offers the opportunity to
analyse the different metal-enrichment events that these systems can
experience. Indeed, 47~Tucanae, besides displaying a well-elongated
1P, shows a very broad ChM sequence of 1P and 2P stars, rather than a
clearly split ChM as in Type~II clusters.  
Its broad ChM suggests that it can include metal enriched stars like
the {\it anomalous} stars in Type~II clusters.  Thus, 47~Tucanae has
not been formally classified as a Type~II GC by \citet{Mil17}, but
represents an ideal target for an in-depth investigation of both the
chemical abundance pattern within 1P stars and other metal enrichment
channels associated with its candidate {\it anomalous} stars which can
the progeny of a faint sub giant branch 
populated by about 10\% of the stars \citep{JA09}. 

The layout of this paper is as follows:
Section~\ref{sec:data} presents the photometric and spectroscopic
data; Section~\ref{sec:atm} describes how we derive atmospheric
parameters and the chemical abundance analysis;
Section~\ref{sec:results} presents our results, that are discussed and
summarised in Sections~\ref{sec:discussion}-\ref{sec:summary}.

%%%%%%%%%%%%%%%%%%%%%%%%%%%%%%%%%%%

\startlongtable
\begin{deluxetable*}{c cc c cc ccc c c}
\tablewidth{10pt}
\tabletypesize{\scriptsize}
\tablecaption{Coordinates, Photometric Information (including the location on the ChM), Radial Velocities, with associated rms from the \# (number) exposures and maximum time between observations (Time interval) for our 29 spectroscopic targets observed with UVES. The {\it anomalous} candidate stars have been highlighted in the last column.}\label{tab:targets}
\tablehead{
\colhead{ID} & RA  & DEC          &  \colhead{$V$} & \colhead{\x} & \colhead{\y}           & \colhead{RV} & \colhead{rms$_{\mathrm {RV}}$} & \colhead{\#}  & \colhead{Time interval} &{\it anomalous}  \\
             & [J2000] & [J2000]      & \colhead{[mag]} & \colhead{[mag]}&\colhead{[mag]}          & \colhead{[\kmsec]} & \colhead{[\kmsec]} &   & [months]&}
\startdata
  8672       & 00:24:20.5	&  $-$72:06:00.19  & 14.97 & $-$0.2192    &  $+$0.0446   & $-$13.99  & 0.93 &  9      & 14 & no \\
  116624     & 00:23:52.9	&  $-$72:04:18.59  & 13.81 & $-$0.1252    &  $+$0.3932   & $-$17.56  & 0.27 &  9      & 14 & yes   \\
  11746      & 00:24:15.6 &  $-$72:06:04.31  & 14.06 & $-$0.0514    &  $+$0.0653   & $+$3.58   & 0.29 &  9      & 14 & no   \\
  44780      & 00:24:16.1	&  $-$72:05:32.89  & 13.64 & $-$0.1588    &  $+$0.0013   & $-$36.37  & 0.26 &  9      & 14 & no   \\
  68444      & 00:23:58.0 &  $-$72:05:03.00  & 14.90 & $-$0.1876    &  $+$0.0951   & $-$12.37  & 0.35 &  9      & 14 & no  \\
  71504      & 00:23:54.8 &  $-$72:05:32.58  & 14.54 & $-$0.0861    &  $+$0.0603   & $-$27.26  & 0.27 &  9      & 14 & no  \\
  92267      & 00:24:12.6	&  $-$72:04:23.67  & 13.43 & $-$0.3425    &  $+$0.3754   & $-$13.04  & 0.27 &  11     & 14 & no   \\
  115804     & 00:23:53.6	&  $-$72:04:48.31  & 14.00 & $-$0.1349    &  $+$0.0846   & $-$28.39  & 0.08 &  2      & 2  & no  \\
  22386      & 00:24:02.7	&  $-$72:05:46.67  & 14.00 & $+$0.0107    &  $+$0.2369   & $-$19.00  & 0.06 &  2      & 2  & yes  \\
  28653      & 00:23:54.6	&  $-$72:05:46.67  & 14.53 & $-$0.0891    &  $+$0.0871   & $-$16.39  & 0.25 &  10     & 5  & no \\
  41968      & 00:24:19.0	&  $-$72:05:13.88  & 14.33 & $-$0.0929    &  $+$0.1460   & $-$27.99  & 0.10 &  2      & 2  & no  \\   
  61006      & 00:24:03.6	&  $-$72:05:06.60  & 14.52 & $-$0.1427    &  $+$0.0568   & $-$19.85  & 0.10 &  2      & 2  & no  \\   
  73063      & 00:23:53.2	&  $-$72:05:25.31  & 14.44 & $-$0.2656    &  $+$0.4352   & $-$5.87   & 0.06 &  2      & 2  & yes  \\ 
  84553      & 00:24:21.0 &  $-$72:04:49.08  & 14.45 & $-$0.1813    &  $+$0.1326   & $-$33.42  & 0.13 &  2      & 2  & no  \\
  10619      & 00:24:17.1	&  $-$72:06:10.60  & 13.47 & $-$0.0842    &  $-$0.0339   & $+$1.33   & 0.22 &  9      & 13 & no   \\
  112499     & 00:23:57.2	&  $-$72:04:17.30  & 14.13 & $-$0.2180    &  $+$0.0792   & $+$10.52  & 0.28 &  9      & 13 & no   \\   
  143175     & 00:23:59.8	&  $-$72:03:53.10  & 14.58 & $-$0.1573    &  $+$0.0434   & $-$31.93  & 0.28 &  9      & 13 & no   \\
  22746      & 00:24:01.9	&  $-$72:06:05.41  & 13.46 & $+$0.0110    &  $-$0.0008   & $-$22.62  & 0.21 &  9      & 13 & no   \\
  70896      & 00:23:55.5	&  $-$72:05:00.48  & 13.77 & $-$0.0069    &  $+$0.0662   & $-$1.59   & 6.72 &  9      & 13 & no   \\
  83377      & 00:24:22.9	&  $-$72:04:31.58  & 14.44 & $-$0.0451    &  $-$0.0195   & $-$26.26  & 0.21 &  9      & 13 & no   \\
  86528      & 00:24:18.5	&  $-$72:04:16.09  & 14.82 & $-$0.0041    &  $+$0.0420   & $+$4.21   & 0.31 &  9      & 13 & no   \\
  90842      & 00:24:14.0	&  $-$72:04:14.20  & 14.53 & $+$0.0042    &  $+$0.0076   & $+$4.39   & 0.25 &  5      & 13 & no   \\
  146316     & 00:23:55.4 &  $-$72:04:00.79  & 14.67 & $-$0.1399    &  $+$0.0973   & $-$24.38  & 0.24 &  3      & 5  & no \\   
  15613      & 00:24:10.4 &  $-$72:05:58.37  & 13.81 & $-$0.1329    &  $+$0.0712   & $-$9.06   & 0.34 &  8      & 5  & no \\
  69465      & 00:23:56.8 &  $-$72:05:09.60  & 13.88 & $-$0.1467    &  $+$0.1009   & $+$12.42  & 0.29 &  8      & 5  & no \\   
  86281      & 00:24:18.9 &  $-$72:04:35.89  & 13.64 & $-$0.0327    &  $+$0.0670   & $+$4.75   & 0.24 &  8      & 5  & no \\
  91298      & 00:24:13.5 &  $-$72:04:31.50  & 14.36 & $-$0.2290    &  $+$0.3097   & $-$19.00  & 0.33 &  8      & 5  & no  \\
  9946       & 00:24:18.2 &  $-$72:06:03.79  & 14.80 & $-$0.0361    &  $-$0.0199   & $-$14.92  & 0.24 &  8      & 5  & no  \\
  100719     & 00:24:06.2	&  $-$72:04:48.78  & 13.69 & $-$0.2126    &  $+$0.0517   & $-$17.87  & 0.23 &  7      &12  & no\\\hline
\enddata
\end{deluxetable*}

%%%%%%%%%%%%%%%%%%%%%%%%%%%%%%%%%%%%%%%%%%%%%%%%%%%%%%%%%%%%%

\section{Data}\label{sec:data}

\subsection{The chromosome map of 47~Tucanae \label{sec:phot_data}} 

\begin{figure}
   \centering
   \includegraphics[width=0.483\textwidth]{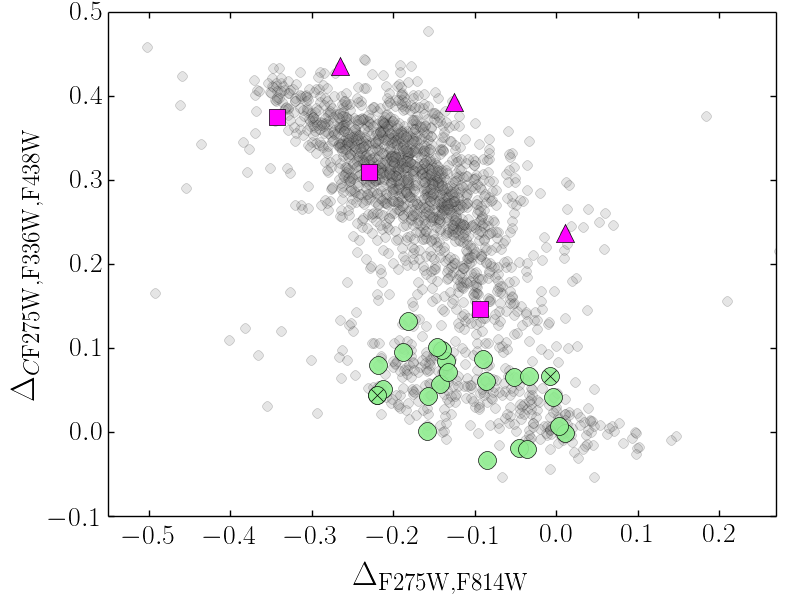}
      \caption{The chromosome map of 47~Tucanae from \citet{Mil17}.
      Grey dots represent the complete sample of RGB stars used to construct the map.
        Our spectroscopic targets observed with UVES are highlighted
        with large coloured symbols. Specifically, green circles are
        1P stars, magenta squares are {\it normal} 2P stars, while
        magenta triangles represent candidate {\it anomalous}
        stars. The locations of two RV binaries are indicated by black
        crosses.}   
        \label{fig:chm}
   \end{figure}

\citet{Mil17} analysed the ChMs of 57 GCs, including
47~Tucanae, whose ChM is shown in Figure~\ref{fig:chm}.
As noticed by \citet{Mil17}, clearly, the ChM of this cluster covers a
large range along the \x\ axis, which is not consistent with the 1P
stars being chemically homogeneous.  

In this work we analyse high-resolution spectra for the targets
shown in Figure~\ref{fig:chm}, and couple our spectroscopic data (see Section~\ref{sec:spec_data}) with high-precision
photometry from the $HST$.
The photometric data come from the $HST$ UV Legacy
Survey that is designed to investigate multiple stellar populations in GCs
\citep[GO-13297,][]{Piotto15}.
Details on the images and on the data reduction can be found in
\citet{Piotto15} and \citet{Mil17}.
Our photometry has been corrected for differential reddening effects as in
\citet{APM12, APM12bin}.

%%_________________________________________________________________________
%%
%%__________________________________________________________________________
%%

\subsection{The spectroscopic dataset}\label{sec:spec_data}

Our spectroscopic data have been acquired using the FLAMES Ultraviolet
and Visual Echelle Spectrograph \citep[FLAMES-UVES,][]{Pasquini00} on
the European Southern Observatory’s (ESO) Very Large 
Telescope (VLT), through the programme 105.20NB. 
The observations were taken in the standard RED580 setup, which has a
wavelength coverage 4726-6835~\AA\ and a resolution of $R \sim$47,000
\citep{Dekker00}. Each target has been selected to pass the isolation
criteria, namely no neighbors brighter than 1.5 mag within two fiber
radii.  

The stars were chosen to span a range in \x\ as large
as possible, and a quite narrow range in \y. Indeed, our primary goal
was to observe red giants associated to the 1P on the ChM.  
As shown in Figure~\ref{fig:chm}, most of our targets lie on the 1P,
spanning a range of $\sim$0.20 in \x.  
Because of the limited sample of isolated targets located within the
inner $\sim$3\arcmin$\times$3\arcmin\ field, where the ChM has been
constructed, not all the UVES fibers could be fed with 1P stars
avoiding fiber collisions. Hence, we had the opportunity to observe a    
few targets on the 2P as well, namely three stars on the blue and 
three other stars on the red edge of the ChM as indicated in
Figure~\ref{fig:chm}. The red targets might be associated with stars
that have an enhanced heavy element composition in Type~II GCs
\citep[e.g.][]{Mil17, Mar19a, Mar21}. 

\begin{figure*}
   \centering
   \includegraphics[width=0.9\textwidth]{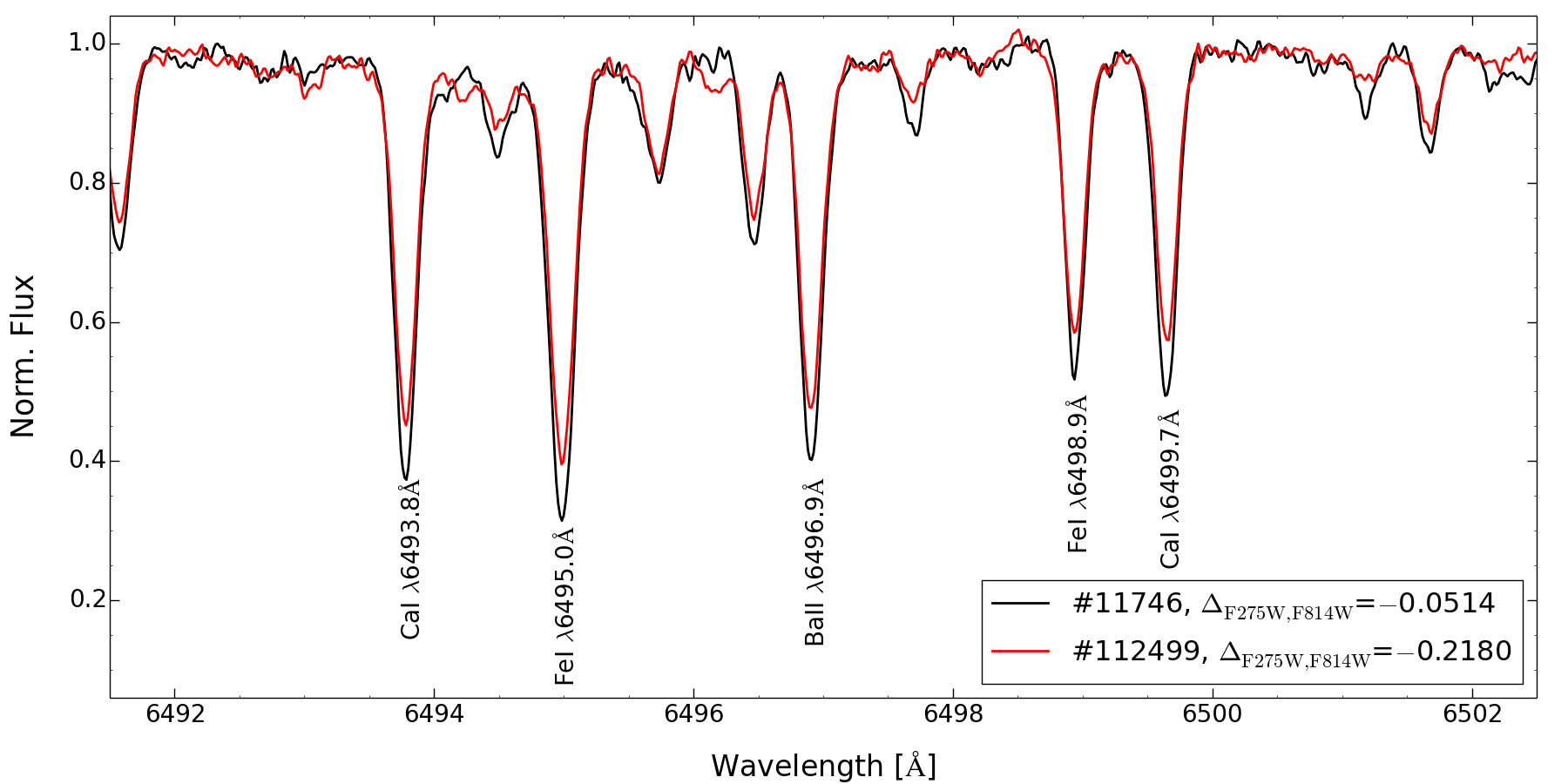}
      \caption{Examples of two portions of FLAMES@UVES spectra
        including some analyzed Fe and Ca spectral features, plus the
        Ba line 6496.9~\AA.  
      The displayed spectra are relative to stars \#11746 and \#112499
      which have similar atmospheric parameters but different \x\
      values. Overall, the spectral features of star \#11746 (with higher
        \x) are deeper.}  
        \label{fig:spectra}
   \end{figure*}

Spectra are based on a number of multiple 2775s exposures, which
varies from a minimum of 2 to a maximum of 11, depending both on the
brightness and priority of each target.  
Individual exposures have been taken a few months apart to allow us to
detect variations in radial velocities (RVs), and identify possible
binaries in our sample. 
Data were reduced using the FLAMES-UVES pipelines within the
EsoReflex interface\footnote{\sf
  {https://www.eso.org/sci/software/esoreflex/}} \citep{Ballester00}, 
including bias subtraction, flat-field correction, wavelength calibration, sky subtraction and spectral rectification. 
As an example, Figure~\ref{fig:spectra} shows two portions of our final spectra.

RVs were derived using the {\sc iraf@FXCOR} task, which
cross-correlates the object spectrum with a template. For the template
we used a synthetic spectrum obtained 
through MOOG\footnote{{\sf http://www.as.utexas.edu/~chris/moog.html}} 
\citep[version June 2014,][]{moog}, computed with a model
stellar atmosphere interpolated from the 
\citet{C&K} grid, adopting parameters
(effective temperature/surface gravity/microturbulence/metallicity) =
(\teff/\logg/\vmicro/[A/H]) = (4500~K/2.0/2.0~\kmsec/$-0.70$~\\dex).  
Each spectrum was corrected to the restframe system, and observed RVs were then
corrected to the heliocentric system. 
Table~\ref{tab:targets} lists the observed targets, together with
their coordinates, photometric information, RVs, number of exposures
and time interval between the observations.

We notice that two stars in our sample, namely \#8672 and \#70896, have larger root mean square deviations (rms) in their RV
values obtained from different exposures (rms$_{\rm RV}$), hence we
treat these two objects as probable binaries, and highlight their
location in the figures with black crosses. 
However, we warn the reader that we cannot rule out the presence of
other binaries with no sizeable differences in the RV values in our
sample.  

The final mean heliocentric RV for our 47~Tucanae giants, excluding
the two binary candidates, is
$\langle$RV$\rangle$=$-14.2\pm2.8$~\kmsec\ ($\sigma$=14.1~\kmsec), 
which lies within $\sim$1~$\sigma$ of the value listed in \citet{BaumgardtHilker18}
of $\langle$RV$\rangle$=$-17.2\pm0.1$~\kmsec  $\sigma$=11.50~\kmsec\
within an average distance of 18.02 arcsec from the cluster centre). 

Finally, the individual exposures for each star have been co-added.
The signal-to-noise (S/N) ratio for the combined spectra 
around $\lambda$6300~\AA\ ranges from S/N$\sim$60 to
$\sim$200, depending on the brightness of the star and the number of exposures.

%%_________________________________________________________________________
%%
%%__________________________________________________________________________
%%

\section{Chemical abundance analysis}\label{sec:atm}

The UVES spectra have a spectral coverage high enough to allow us a
fully-spectroscopical estimate of the stellar parameters, \teff,
\logg, [A/H] and \vmicro. Indeed the number of measured Fe\,I
and Fe\,II lines for each star is typically 120 and 12,
respectively.  
Briefly, \teff\ and \logg\ were inferred by imposing the excitation
potential (E.P.) equilibrium of the Fe\,I lines and the
ionization equilibrium between Fe\,I and Fe\,II spectral
features, respectively. For \logg\ we further imposed Fe\,II
abundances 0.05-0.07~dex higher than the Fe\,I ones to adjust
for non-local thermodynamic equilibrium (non-LTE) effects
\citep{Bergemann12, Lind12}. \vmicro\ was set to minimize any
dependence of Fe\,I abundances as a function of equivalent width
(EW).   
The adopted atmospheric parameters are listed in Table~\ref{tab:atm}.

As our sample of stars is composed of giants with information on the
ChM obtained from $HST$ filters, our targets are located in the
2.7\arcmin$\times$2.7\arcmin\ innermost region of 47~Tucanae.  
Hence, we have only star \#69465 in common with \citet{carretta09a}.
For this target, our inferred atmospheric parameters are in excellent
agreement with those of Carretta and collaborators. Indeed, we obtain
differences (Carretta minus this work) of $\Delta\teff=+27$~K,
$\Delta\logg=-0.01$, and $\Delta\vmicro=-0.02$~\kmsec. The larger
difference we obtain is in the Fe abundance,  
$\Delta$[Fe/H]=$+0.13$~dex, which could be likely explained with the
different lines used in the two studies, as we exploit UVES data with
a larger wavelength coverage and a higher resolution than the GIRAFFE
spectra analyzed by \citet{carretta09a}. 
 
Following our previous work based on data with similar quality, we
here adopt typical internal uncertainties in the atmospheric
parameters of 50~K for \teff, 0.20~dex for \logg, 0.20~\kmsec\ for
\vmicro, and 0.10~dex for metallicity \citep[see][]{Mar19b}. We also
emphasize that systematic errors might be larger.

%%%%%%%%%%%%%%%%%%%%%%%%%%%%%%%%%%%%%%%%%%%%%%%%%%%%%%%%%%%%%%%%%%%%%%%%%%%%%%%%%%%%%%%%%%%%%%%%%%%% 
\begin{table}
\centering
\caption{Adopted atmospheric parameters.}\label{tab:atm}
\begin{tabular}{c c c c c}
\hline\hline
ID           &  \teff & \logg  & [Fe/H] & \vmicro \\   
             & (K)    & (cgs)  &   dex  & (\kmsec)   \\  \hline
8672   &4600 &2.20 &$-$0.87 &1.10\\
116624 &4580 &2.20 &$-$0.70 &1.30\\
11746  &4640 &2.30 &$-$0.76 &1.27\\
44780  &4550 &2.15 &$-$0.83 &1.33\\
68444  &4970 &3.05 &$-$0.83 &0.65\\
71504  &4800 &2.70 &$-$0.80 &0.78\\
92267  &4530 &2.07 &$-$0.82 &1.19\\
115804 &4660 &2.33 &$-$0.76 &1.26\\
22386  &4660 &2.35 &$-$0.65 &1.33\\
28653  &4830 &2.75 &$-$0.72 &1.12\\
41968  &4780 &2.60 &$-$0.74 &1.17\\
61006  &4900 &3.00 &$-$0.79 &0.81\\
73063  &4880 &2.80 &$-$0.66 &1.15\\
84553  &4700 &2.30 &$-$0.95 &1.12\\
10619  &4470 &1.95 &$-$0.85 &1.47\\
112499 &4700 &2.47 &$-$0.88 &0.90\\
143175 &4830 &2.80 &$-$0.79 &1.10\\
22746  &4450 &1.97 &$-$0.79 &1.47\\
70896  &4550 &2.15 &$-$0.77 &1.30\\
83377  &4780 &2.60 &$-$0.74 &1.18\\
86528  &4880 &2.90 &$-$0.74 &0.84\\
90842  &4760 &2.60 &$-$0.75 &0.95\\
146316 &4920 &2.95 &$-$0.68 &0.90\\
15613  &4600 &2.20 &$-$0.82 &1.41\\
69465  &4550 &2.12 &$-$0.87 &1.15\\
86281  &4500 &2.10 &$-$0.82 &1.40\\
91298  &4870 &3.02 &$-$0.73 &1.07\\
9946   &4900 &2.85 &$-$0.67 &1.12\\
100719 &4650 &2.30 &$-$0.97 &0.93 \\\hline
\end{tabular}
\end{table}
%%%%%%%%%%%%%%%%%%%%%%%%%%%%%%%%%%%%%%%%%%%%%%%%%%%%%%%%%%%%%%%%%%%%%%%%%%%%%%%%%%%%%%%%%%%%%%%%%%%% 

Chemical abundances have been inferred for 22 elements, namely 
Na, Mg, Al, Si, Ca, Sc ({\sc ii}), Ti ({\sc i} and {\sc ii}), V, Cr
({\sc i} and {\sc ii}), Mn, Fe ({\sc i} and {\sc ii}), Co, Ni, Cu, Zn,
Y ({\sc ii}), Zr ({\sc ii}), Ba ({\sc ii}), La ({\sc ii}), Pr ({\sc
  ii}), Nd ({\sc ii}), Eu ({\sc ii}).  
The abundance analysis has been performed by employing  the MOOG code
\citep{moog} with the alpha-enhanced Kurucz model atmospheres of
\citet{C&K}.  
Chemical abundances have been inferred
by using an EW-based analysis; when this was not possible because of
blends or hyperfine/isotopic splitting we performed spectral
synthesis. We now comment on some of the transitions that we used.

{\it Proton-capture elements:}
Sodium abundances were determined from the EWs of the Na\,{\sc i} doublets at
$\sim$5680~\AA\ and $\sim$6150~\AA, aluminum from the synthesis of the doublet at
$\sim$6667~\AA, and magnesium from the EWs of the
transitions at $\sim$5528, 5711~\AA. 
Because of the low radial velocity of the cluster, no reliable
abundances could be inferred for oxygen, as the [O I] line was masked
by much stronger telluric features.  

{\it Manganese:}
Spectral lines around
5399, 5420, 5433, 6014, 6022~\AA\ have been analyzed to infer chemical
abundances for Mn. 
In our synthesis we assume f(${\phantom{}}^{55}$Mn)=1.00 
and hyperfine splitting from \citet{LawlerA, LawlerB} and \citet{Kur09}.

{\it Copper:}
For Cu, we synthesised the spectral feature at $\sim$5105~\AA, with
both hyperfine and isotopic splitting from \citet{Kur09}. Solar-system isotopic fractions
of f(${\phantom{}}^{63}$Cu)=0.69 and
f(${\phantom{}}^{65}$Cu)=0.31 have been assumed.  

{\it Neutron-capture elements:} 
We inferred abundances for Y, Zr, Ba, La, Pr, Nd, and Eu.
A spectral synthesis analysis has been employed for Zr
($\lambda$5112~\AA), Ba ($\lambda$5854, 6497~\AA), La
($\lambda$4921, 5115, 5291, 5302, 5303, 6262, 5936, 6390, 6774~\AA), Pr
($\lambda$5323~\AA), and Eu ($\lambda$6645~\AA), affected by
hyperfine and/or isotopic splitting and/or blending features.
Barium abundances were computed assuming the \citet{McW98} 
$r$-process isotopic composition and hyperfine splitting. 

%%%%%%%%%%%%%%%%%%%%%%%%%%%%%%%%%%%%%%%%%%%%%%%%%%%%%%%%%%%%%%%%%%%%%%%%%%%%%%%%%%%%%%%%%%%%%%%%%%
\startlongtable
\begin{deluxetable*}{ccccccccccccccccccccccc}
\tablewidth{6pt}
\tabletypesize{\scriptsize}
\tablecaption{Analyzed Chemical Abundances from Na to Sc. Average values for 1P and 2P stars are given for each element at the bottom of the table.\label{tab:NaToSc}}
\tablehead{
STAR &   [Na/Fe] & $\sigma$ & \# & [Mg/Fe] & $\sigma$ & \# & [Al/Fe] & $\sigma$ & \# & [Si/Fe] & $\sigma$ & \# & [Ca/Fe] & $\sigma$ & \# & [Sc/Fe] & $\sigma$ & \# 
}
\startdata
8672    & 0.37 & 0.07 & 3 & 0.25 & 0.00 & 1 & 0.43 & 0.05 & 2 & 0.43 & 0.12 & 8 & 0.40 & 0.27 & 18 & 0.13 & 0.25 & 3     \\
116624  & 0.64 & 0.06 & 4 & 0.35 & 0.00 & 1 & 0.47 & 0.12 & 2 & 0.49 & 0.08 & 8 & 0.32 & 0.20 & 17 & 0.22 & 0.14 & 6     \\
11746   & 0.38 & 0.09 & 4 & 0.45 & 0.00 & 1 & 0.21 & 0.04 & 2 & 0.44 & 0.05 & 7 & 0.30 & 0.15 & 20 & 0.24 & 0.14 & 6     \\
44780   & 0.24 & 0.10 & 4 & 0.46 & 0.00 & 1 & 0.21 & 0.06 & 2 & 0.44 & 0.06 & 7 & 0.30 & 0.17 & 19 & 0.23 & 0.12 & 6     \\
68444   & 0.23 & 0.05 & 4 & 0.26 & 0.00 & 1 & 0.06 & 0.10 & 2 & 0.34 & 0.09 & 7 & 0.28 & 0.14 & 17 & 0.18 & 0.11 & 6     \\
71504   & 0.18 & 0.06 & 4 & 0.29 & 0.00 & 1 & 0.15 & 0.05 & 2 & 0.35 & 0.02 & 6 & 0.26 & 0.13 & 19 & 0.23 & 0.15 & 6     \\
92267   & 0.75 & 0.06 & 4 & 0.43 & 0.00 & 1 & 0.32 & 0.03 & 2 & 0.44 & 0.06 & 7 & 0.32 & 0.19 & 17 & 0.27 & 0.15 & 6     \\
115804  & 0.18 & 0.13 & 4 & 0.41 & 0.00 & 1 & 0.16 & 0.04 & 2 & 0.41 & 0.07 & 8 & 0.30 & 0.16 & 19 & 0.20 & 0.13 & 6     \\
22386   & 0.62 & 0.13 & 4 & 0.51 & 0.00 & 1 & 0.28 & 0.01 & 2 & 0.41 & 0.09 & 8 & 0.33 & 0.17 & 18 & 0.23 & 0.09 & 6     \\
28653   & 0.26 & 0.12 & 4 & 0.41 & 0.00 & 1 & 0.12 & 0.05 & 2 & 0.37 & 0.07 & 7 & 0.26 & 0.13 & 20 & 0.24 & 0.15 & 6     \\
41968   & 0.36 & 0.06 & 4 & 0.41 & 0.00 & 1 & 0.15 & 0.02 & 2 & 0.38 & 0.10 & 8 & 0.24 & 0.13 & 18 & 0.22 & 0.11 & 6     \\
61006   & 0.20 & 0.05 & 4 & 0.28 & 0.00 & 1 & 0.07 & 0.06 & 2 & 0.39 & 0.10 & 6 & 0.22 & 0.12 & 16 & 0.26 & 0.21 & 5     \\
73063   & 0.50 & 0.05 & 4 & 0.37 & 0.00 & 1 & 0.23 & 0.10 & 2 & 0.35 & 0.09 & 7 & 0.25 & 0.17 & 19 & 0.30 & 0.13 & 6     \\
84553   & 0.47 & 0.14 & 3 & 0.40 & 0.00 & 1 & 0.21 & 0.07 & 2 & 0.39 & 0.09 & 7 & 0.30 & 0.09 & 10 & 0.12 & 0.13 & 6     \\
10619   & 0.29 & 0.12 & 4 & 0.51 & 0.00 & 1 & 0.24 & 0.08 & 2 & 0.45 & 0.04 & 8 & 0.28 & 0.16 & 17 & 0.22 & 0.11 & 6     \\
112499  & 0.20 & 0.04 & 4 & 0.35 & 0.00 & 1 & 0.10 & 0.04 & 2 & 0.38 & 0.06 & 6 & 0.23 & 0.12 & 16 & 0.26 & 0.15 & 6     \\
143175  & 0.20 & 0.11 & 4 & 0.38 & 0.00 & 1 & 0.11 & 0.05 & 2 & 0.39 & 0.06 & 8 & 0.26 & 0.15 & 18 & 0.28 & 0.12 & 6     \\
22746   & 0.30 & 0.14 & 4 & 0.47 & 0.00 & 1 & 0.20 & 0.06 & 2 & 0.45 & 0.04 & 8 & 0.28 & 0.17 & 19 & 0.22 & 0.11 & 6     \\
70896   & 0.27 & 0.09 & 4 & 0.45 & 0.00 & 1 & 0.14 & 0.02 & 2 & 0.43 & 0.02 & 7 & 0.31 & 0.15 & 17 & 0.22 & 0.12 & 6     \\
83377   & 0.25 & 0.12 & 4 & 0.46 & 0.00 & 1 & 0.16 & 0.04 & 2 & 0.37 & 0.03 & 7 & 0.28 & 0.14 & 19 & 0.22 & 0.12 & 6     \\
86528   & 0.19 & 0.07 & 4 & 0.34 & 0.00 & 1 & 0.18 & 0.05 & 2 & 0.39 & 0.07 & 8 & 0.27 & 0.19 & 16 & 0.28 & 0.13 & 6     \\
90842   & 0.18 & 0.06 & 4 & 0.40 & 0.00 & 1 & 0.12 & 0.03 & 2 & 0.37 & 0.17 & 7 & 0.30 & 0.17 & 18 & 0.28 & 0.18 & 5     \\
146316  & 0.18 & 0.06 & 4 & 0.16 & 0.20 & 2 & 0.06 & 0.12 & 2 & 0.28 & 0.21 & 7 & 0.26 & 0.15 & 18 & 0.30 & 0.08 & 5     \\
15613   & 0.38 & 0.10 & 4 & 0.48 & 0.00 & 1 & 0.18 & 0.04 & 2 & 0.46 & 0.07 & 8 & 0.29 & 0.15 & 17 & 0.22 & 0.12 & 6     \\
69465   & 0.28 & 0.08 & 4 & 0.39 & 0.00 & 1 & 0.14 & 0.02 & 2 & 0.46 & 0.08 & 7 & 0.30 & 0.16 & 19 & 0.24 & 0.14 & 6     \\
86281   & 0.38 & 0.11 & 4 & 0.46 & 0.00 & 1 & 0.18 & 0.07 & 2 & 0.49 & 0.05 & 7 & 0.28 & 0.16 & 18 & 0.26 & 0.11 & 6     \\
91298   & 0.56 & 0.04 & 4 & 0.31 & 0.00 & 1 & 0.19 & 0.12 & 2 & 0.42 & 0.07 & 7 & 0.27 & 0.15 & 15 & 0.31 & 0.15 & 6     \\
9946    & 0.17 & 0.12 & 4 & 0.37 & 0.00 & 1 & 0.14 & 0.03 & 2 & 0.32 & 0.06 & 8 & 0.26 & 0.12 & 18 & 0.25 & 0.08 & 6     \\
100719  & 0.30 & 0.08 & 4 & 0.37 & 0.00 & 1 & 0.15 & 0.05 & 2 & 0.39 & 0.11 & 8 & 0.34 & 0.17 & 16 & 0.15 & 0.15 & 6     \\\hline
$<$[X/Fe]$>_{\rm 1P}$&0.26&&&0.38 &      &   & 0.16 &      &   & 0.40 &      &   & 0.28 &      &    & 0.23 &      & \\
$\pm$               &0.02&&&0.02 &      &   & 0.02 &      &   & 0.01 &      &   & 0.01 &      &    & 0.01 &      & \\
$\sigma$            &0.08&&&0.09 &      &   & 0.08 &      &   & 0.05 &      &   & 0.04 &      &    & 0.04 &      & \\
$<$[X/Fe]$>_{\rm 2P}$&0.57&&&0.40 &      &   & 0.27 &      &   & 0.41 &      &   & 0.29 &      &    & 0.24 &      & \\
$\pm$               &0.06&&&0.03 &      &   & 0.05 &      &   & 0.02 &      &   & 0.01 &      &    & 0.03 &      & \\
$\sigma$            &0.13&&&0.07 &      &   & 0.11 &      &   & 0.05 &      &   & 0.04 &      &    & 0.06 &      & \\\hline
\enddata
\end{deluxetable*}
%%%%%%%%%%%%%%%%%%%%%%%%%%%%%%%%%%%%%%%%%%%%%%%%%%%%%%%%%%%%%%%%%%%%%%%%%%%%%%%%%%%%%%%%%%%%%%%%%%%% 

%%%%%%%%%%%%%%%%%%%%%%%%%%%%%%%%%%%%%%%%%%%%%%%%%%%%%%%%%%%%%%%%%%%%%%%%%%%%%%%%%%%%%%%%%%%%%%%%%%
\startlongtable
\begin{deluxetable*}{c ccc ccc ccc ccc ccc ccc ccc ccc}
\tabletypesize{\tiny}
\tablecaption{Analyzed Chemical Abundances from Ti to Ni. Average values for 1P and 2P stars are given for each element at the bottom of the table.\label{tab:TiToNi}}
\tablehead{
STAR &[Ti/Fe]{\sc i} & $\sigma$ & \# & [Ti/Fe]{\sc ii} & $\sigma$ & \# & [V/Fe]  & $\sigma$ & \# & [Cr/Fe]{\sc i} & $\sigma$ & \# & [Cr/Fe]{\sc ii} & $\sigma$ & \# & [Mn/Fe] & $\sigma$ & \# & [Co/Fe] & $\sigma$ & \# & [Ni/Fe] & $\sigma$ & \# 
}
\startdata
8672    & 0.35 & 0.23 & 14 & 0.30 & 0.20 & 3 & 0.22 & 0.16& 11 & $-$0.31 & 0.07 & 4 & $-$0.15 & 0.24 & 2 & $-$0.55 &0.17 & 5 & $-$0.18 & 0.06 & 2 & 0.13 & 0.23 & 24 \\
116624  & 0.33 & 0.15 & 21 & 0.25 & 0.06 & 4 & 0.34 & 0.13& 12 & $-$0.15 & 0.10 & 4 & $+$0.05 & 0.01 & 2 & $-$0.26 &0.02 & 5 & $+$0.07 & 0.10 & 2 & 0.10 & 0.16 & 22 \\
11746   & 0.33 & 0.10 & 23 & 0.31 & 0.05 & 4 & 0.33 & 0.21& 14 & $-$0.10 & 0.06 & 4 & $+$0.20 & 0.05 & 2 & $-$0.23 &0.04 & 5 & $+$0.08 & 0.12 & 2 & 0.11 & 0.15 & 24 \\
44780   & 0.33 & 0.14 & 23 & 0.38 & 0.07 & 6 & 0.45 & 0.25& 16 & $-$0.07 & 0.06 & 5 & $+$0.10 & 0.07 & 2 & $-$0.24 &0.02 & 5 & $+$0.10 & 0.14 & 2 & 0.11 & 0.14 & 27 \\
68444   & 0.31 & 0.10 & 22 & 0.34 & 0.07 & 4 & 0.25 & 0.12& 13 & $-$0.10 & 0.13 & 4 & $+$0.08 & 0.07 & 2 & $-$0.31 &0.05 & 5 & $-$0.04 & 0.10 & 2 & 0.10 & 0.13 & 25 \\
71504   & 0.31 & 0.11 & 22 & 0.28 & 0.06 & 4 & 0.26 & 0.14& 13 & $-$0.06 & 0.10 & 4 & $+$0.01 & 0.05 & 2 & $-$0.33 &0.04 & 5 & $-$0.01 & 0.11 & 2 & 0.10 & 0.15 & 25 \\
92267   & 0.33 & 0.14 & 23 & 0.29 & 0.09 & 4 & 0.39 & 0.22& 14 & $-$0.12 & 0.10 & 4 & $+$0.05 & 0.03 & 2 & $-$0.31 &0.03 & 5 & $+$0.05 & 0.17 & 2 & 0.11 & 0.15 & 24 \\
115804  & 0.30 & 0.11 & 22 & 0.27 & 0.03 & 4 & 0.30 & 0.22& 13 & $-$0.10 & 0.12 & 4 & $+$0.06 & 0.01 & 2 & $-$0.32 &0.03 & 5 & $+$0.06 & 0.13 & 2 & 0.08 & 0.14 & 25 \\
22386   & 0.35 & 0.12 & 21 & 0.27 & 0.09 & 4 & 0.46 & 0.21& 13 & $-$0.06 & 0.10 & 4 & $+$0.05 & 0.28 & 2 & $-$0.27 &0.04 & 5 & $+$0.12 & 0.20 & 2 & 0.06 & 0.15 & 23 \\
28653   & 0.31 & 0.14 & 23 & 0.31 & 0.04 & 4 & 0.26 & 0.15& 14 & $-$0.04 & 0.10 & 3 & $+$0.09 & 0.15 & 2 & $-$0.27 &0.03 & 5 & $+$0.08 & 0.06 & 2 & 0.11 & 0.12 & 23 \\
41968   & 0.35 & 0.11 & 23 & 0.28 & 0.08 & 4 & 0.31 & 0.16& 13 & $-$0.06 & 0.11 & 4 & $+$0.07 & 0.02 & 2 & $-$0.29 &0.03 & 5 & $+$0.07 & 0.15 & 2 & 0.09 & 0.14 & 23 \\
61006   & 0.30 & 0.13 & 20 & 0.30 & 0.14 & 4 & 0.22 & 0.14& 13 & $-$0.07 & 0.20 & 4 & $+$0.17 & 0.15 & 2 & $-$0.38 &0.08 & 5 & $+$0.15 & 0.06 & 2 & 0.08 & 0.14 & 23 \\
73063   & 0.33 & 0.12 & 22 & 0.34 & 0.03 & 4 & 0.38 & 0.25& 12 & $-$0.13 & 0.19 & 4 & $+$0.14 & 0.18 & 2 & $-$0.25 &0.03 & 5 & $+$0.08 & 0.07 & 2 & 0.13 & 0.13 & 23 \\
84553   & 0.44 & 0.17 &  9 & 0.31 & 0.04 & 3 & 0.61 & 0.44&  5 & $+$0.16 & 0.18 & 3 &  --     & --   & 0 & $-$0.32 &0.10 & 5 & $+$0.16 & 0.08 & 2 & 0.09 & 0.16 & 20 \\
10619   & 0.36 & 0.10 & 19 & 0.29 & 0.03 & 3 & 0.39 & 0.26& 14 & $-$0.13 & 0.06 & 4 & $+$0.12 & 0.05 & 2 & $-$0.23 &0.02 & 5 & $+$0.15 & 0.13 & 2 & 0.10 & 0.12 & 25 \\
112499  & 0.31 & 0.07 & 20 & 0.41 & 0.19 & 3 & 0.26 & 0.15& 13 & $-$0.02 & 0.11 & 4 & $+$0.02 & 0.02 & 2 & $-$0.35 &0.05 & 5 & $-$0.02 & 0.08 & 2 & 0.13 & 0.17 & 25 \\
143175  & 0.29 & 0.10 & 23 & 0.35 & 0.05 & 4 & 0.23 & 0.14& 13 & $-$0.11 & 0.10 & 4 & $+$0.10 & 0.06 & 2 & $-$0.28 &0.04 & 5 & $+$0.03 & 0.11 & 2 & 0.11 & 0.12 & 25 \\
22746   & 0.35 & 0.13 & 22 & 0.30 & 0.04 & 4 & 0.46 & 0.26& 12 & $-$0.10 & 0.08 & 4 & $+$0.13 & 0.09 & 2 & $-$0.18 &0.05 & 5 & $+$0.19 & 0.21 & 2 & 0.11 & 0.13 & 25 \\
70896   & 0.34 & 0.14 & 19 & 0.28 & 0.09 & 3 & 0.40 & 0.23& 13 & $-$0.12 & 0.09 & 4 & $+$0.07 & 0.08 & 2 & $-$0.34 &0.07 & 5 & $+$0.19 & 0.19 & 2 & 0.12 & 0.14 & 25 \\
83377   & 0.31 & 0.10 & 23 & 0.26 & 0.02 & 4 & 0.25 & 0.14& 13 & $-$0.04 & 0.10 & 4 & $+$0.12 & 0.14 & 2 & $-$0.25 &0.03 & 5 & $+$0.05 & 0.13 & 2 & 0.09 & 0.13 & 25 \\
86528   & 0.31 & 0.12 & 18 & 0.34 & 0.17 & 2 & 0.22 & 0.17& 12 & $-$0.16 & 0.15 & 3 & $-$0.06 & 0.01 & 2 & $-$0.27 &0.05 & 5 & $+$0.07 & 0.07 & 2 & 0.14 & 0.17 & 24 \\
90842   & 0.28 & 0.11 & 22 & 0.30 & 0.06 & 4 & 0.24 & 0.17& 12 & $-$0.11 & 0.14 & 4 & $+$0.05 & 0.02 & 2 & $-$0.31 &0.04 & 5 & $+$0.09 & 0.14 & 2 & 0.11 & 0.18 & 25 \\
146316  & 0.29 & 0.16 & 23 & 0.30 & 0.04 & 3 & 0.28 & 0.17& 15 & $-$0.10 & 0.22 & 4 & $+$0.22 & 0.26 & 2 & $-$0.28 &0.05 & 5 & $+$0.15 & 0.21 & 2 & 0.14 & 0.16 & 23 \\
15613   & 0.36 & 0.09 & 21 & 0.33 & 0.07 & 4 & 0.35 & 0.19& 13 & $-$0.10 & 0.08 & 3 & $+$0.10 & 0.05 & 2 & $-$0.21 &0.05 & 5 & $+$0.09 & 0.10 & 2 & 0.10 & 0.13 & 24 \\
69465   & 0.28 & 0.10 & 23 & 0.37 & 0.08 & 4 & 0.27 & 0.18& 13 & $-$0.12 & 0.09 & 4 & $+$0.05 & 0.07 & 2 & $-$0.30 &0.02 & 5 & $+$0.05 & 0.11 & 2 & 0.11 & 0.15 & 25 \\
86281   & 0.35 & 0.11 & 20 & 0.33 & 0.04 & 3 & 0.41 & 0.25& 11 & $-$0.09 & 0.06 & 4 & $+$0.10 & 0.11 & 2 & $-$0.21 &0.03 & 5 & $+$0.14 & 0.16 & 2 & 0.14 & 0.16 & 23 \\
91298   & 0.35 & 0.12 & 23 & 0.35 & 0.10 & 4 & 0.30 & 0.14& 13 & $-$0.10 & 0.08 & 4 & $+$0.16 & 0.02 & 2 & $-$0.25 &0.04 & 5 & $+$0.10 & 0.10 & 2 & 0.14 & 0.14 & 24 \\
9946    & 0.33 & 0.15 & 23 & 0.28 & 0.06 & 4 & 0.27 & 0.20& 13 & $+$0.02 & 0.11 & 3 & $+$0.11 & 0.08 & 2 & $-$0.25 &0.01 & 5 & $+$0.09 & 0.05 & 2 & 0.10 & 0.13 & 24 \\
100719  & 0.32 & 0.15 & 22 & 0.26 & 0.05 & 3 & 0.28 & 0.12& 13 & $-$0.10 & 0.08 & 4 & $-$0.03 & 0.02 & 2 & $-$0.38 &0.09 & 5 & $+$0.01 & 0.07 & 2 & 0.10 & 0.13 & 26 \\\hline
$<$[X/Fe]$>_{\rm 1P}$&0.32&&& 0.31 &      &   & 0.30 &     &    & $-$0.10 &      &   & 0.08    &      &   & $-$0.29 &     &   & 0.07    &      &   & 0.11 &      & \\
$\pm$               &0.01&&& 0.01 &      &   & 0.02 &     &    & 0.01    &      &   & 0.02    &      &   & 0.02    &     &   & 0.02    &      &   & 0.00 &      & \\
$\sigma$            &0.02&&& 0.04 &      &   & 0.08 &     &    & 0.06    &      &   & 0.08    &      &   & 0.08    &     &   & 0.08    &      &   & 0.02 &      & \\
$<$[X/Fe]$>_{\rm 2P}$&0.35&&& 0.30 &      &   & 0.40 &     &    & $-$0.07 &      &   & 0.09    &      &   & $-$0.28 &     &   & 0.09    &      &   & 0.10 &      & \\
$\pm$               &0.02&&& 0.01 &      &   & 0.04 &     &    & 0.04    &      &   & 0.02    &      &   & 0.01    &     &   & 0.02    &      &   & 0.01 &      & \\
$\sigma$            &0.04&&& 0.04 &      &   & 0.11 &     &    & 0.11    &      &   & 0.05    &      &   & 0.03    &     &   & 0.04    &      &   & 0.03 &      & \\\hline
\enddata
\end{deluxetable*}
%%%%%%%%%%%%%%%%%%%%%%%%%%%%%%%%%%%%%%%%%%%%%%%%%%%%%%%%%%%%%%%%%%%%%%%%%%%%%%%%%%%%%%%%%%%%%%%%%%%% 

%%%%%%%%%%%%%%%%%%%%%%%%%%%%%%%%%%%%%%%%%%%%%%%%%%%%%%%%%%%%%%%%%%%%%%%%%%%%%%%%%%%%%%%%%%%%%%%%%%
\startlongtable
\begin{deluxetable*}{c c ccc ccc c ccc ccc c ccc c}
\tablewidth{6pt}
\tabletypesize{\tiny}
\tablecaption{Analyzed Chemical Abundances from Cu to Eu. Average values for 1P and 2P stars are given for each element at the bottom of the table.\label{tab:CuToEu}}
\tablehead{
STAR& [Cu/Fe] & [Zn/Fe] &$\sigma$&\#& [Y/Fe] &$\sigma$&\#& [Zr/Fe] &[Ba/Fe]&$\sigma$&\# &[La/Fe]& $\sigma$&\# &[Pr/Fe]&[Nd/Fe]&$\sigma$&\#&[Eu/Fe] }
\startdata
8672    & $-$0.63 & $+$0.03 & 0.00 & 1 & $-$0.14 & 0.26 & 2 & 0.20 & $+$0.40 &  0.04 & 2 & 0.17 & 0.19 & 7 & 0.22 & 0.18 &   -- & 1& 0.40 \\
116624  & $-$0.05 &      -- &   -- & 0 & $+$0.15 & 0.27 & 3 & 0.33 & $+$0.33 &  0.05 & 2 & 0.16 & 0.18 & 9 & 0.12 & 0.30 & 0.15 & 2& 0.46 \\
11746   & $-$0.06 & $+$0.09 & 0.00 & 1 & $+$0.09 & 0.11 & 3 & 0.19 & $+$0.22 &  0.13 & 2 & 0.19 & 0.13 & 9 & 0.29 & 0.30 & 0.16 & 2& 0.50 \\
44780   & $-$0.09 & $+$0.28 & 0.01 & 2 & $+$0.06 & 0.26 & 3 & 0.25 & $+$0.24 &  0.05 & 2 & 0.21 & 0.18 & 9 & 0.27 & 0.38 & 0.15 & 2& 0.54 \\
68444   & $-$0.35 & $+$0.12 & 0.00 & 1 & $+$0.04 & 0.15 & 3 & 0.22 & $+$0.31 &  0.01 & 2 & 0.29 & 0.14 & 7 & 0.30 & 0.33 & 0.20 & 2& 0.62 \\
71504   & $-$0.26 & $+$0.22 & 0.00 & 1 & $+$0.06 & 0.18 & 3 & 0.30 & $+$0.34 &  0.02 & 2 & 0.33 & 0.17 & 8 & 0.33 & 0.41 & 0.16 & 2& 0.56 \\
92267   & $-$0.22 & $+$0.14 & 0.00 & 1 & $+$0.05 & 0.18 & 3 & 0.18 & $+$0.16 &  0.03 & 2 & 0.17 & 0.16 & 9 & 0.23 & 0.36 & 0.13 & 2& 0.44 \\
115804  & $-$0.10 &     --  &  --  & 0 & $+$0.14 & 0.24 & 3 & 0.20 & $+$0.23 &  0.05 & 2 & 0.23 & 0.12 & 9 & 0.15 & 0.47 &   -- & 1& 0.48 \\
22386   & $+$0.00 & $+$0.15 & 0.00 & 1 & $+$0.16 & 0.31 & 3 & 0.32 & $+$0.30 &  0.08 & 2 & 0.34 & 0.06 & 8 & 0.34 & 0.51 &   -- & 1& 0.55 \\
28653   & $-$0.23 & $+$0.20 & 0.00 & 1 & $+$0.09 & 0.14 & 3 & 0.23 & $+$0.32 &  0.11 & 2 & 0.31 & 0.10 & 8 & 0.35 & 0.42 & 0.03 & 2& 0.65 \\
41968   & $+$0.13 & $+$0.25 & 0.00 & 1 & $+$0.15 & 0.31 & 3 & 0.40 & $+$0.15 &  0.21 & 2 & 0.30 & 0.09 & 8 & 0.38 & 0.53 & 0.14 & 2& 0.60 \\
61006   & $-$0.43 & $+$0.24 & 0.00 & 1 & $+$0.09 & 0.13 & 3 & 0.32 & $+$0.24 &  0.19 & 2 & 0.44 & 0.07 & 4 & 0.35 & 0.53 & 0.00 & 2& 9999 \\
73063   & $-$0.05 & $+$0.31 & 0.00 & 1 & $+$0.19 & 0.15 & 3 & 0.26 & $+$0.28 &  0.13 & 2 & 0.27 & 0.14 & 6 & 0.30 & 0.45 & 0.05 & 2& 0.50 \\
84553   & $-$0.31 & $-$0.09 & 0.00 & 1 & $+$0.40 & 0.24 & 3 & 0.24 & $+$0.23 &  0.54 & 2 & 0.41 & 0.38 & 5 & 0.33 & 0.22 &   -- & 1& 9999 \\
10619   & $-$0.01 & $+$0.13 & 0.00 & 1 & $+$0.04 & 0.24 & 3 & 0.23 & $+$0.16 &  0.02 & 2 & 0.19 & 0.20 & 9 & 0.27 & 0.50 &   -- & 1& 0.56 \\
112499  & $-$0.36 & $+$0.23 & 0.00 & 1 & $+$0.06 & 0.25 & 3 & 0.22 & $+$0.30 &  0.13 & 2 & 0.24 & 0.15 & 8 & 0.28 & 0.35 & 0.18 & 2& 0.58 \\
143175  & $+$0.03 &      -- &   -- & 0 & $+$0.17 & 0.11 & 3 & 0.20 & $+$0.31 &  0.18 & 2 & 0.30 & 0.18 & 8 & 0.37 & 0.33 & 0.18 & 2& 0.59 \\
22746   & $+$0.02 &      -- &   -- & 0 & $+$0.09 & 0.23 & 3 & 0.21 & $+$0.16 &  0.00 & 2 & 0.24 & 0.18 & 9 & 0.26 & 0.62 &   -- & 1& 0.53 \\
70896   & $-$0.42 & $+$0.02 & 0.00 & 1 & $-$0.09 & 0.29 & 2 & 0.20 & $-$0.05 &  0.05 & 2 & 0.16 & 0.16 & 9 & 0.25 & 0.39 & 0.17 & 2& 0.54 \\
83377   & $-$0.11 & $+$0.06 & 0.00 & 1 & $+$0.04 & 0.17 & 3 & 0.19 & $+$0.23 &  0.10 & 2 & 0.24 & 0.14 & 9 & 0.20 & 0.35 & 0.20 & 2& 0.65 \\
86528   & $-$0.50 & $-$0.21 & 0.00 & 1 & $-$0.04 & 0.09 & 3 & 0.35 & $+$0.38 &  0.10 & 2 & 0.31 & 0.17 & 7 & 0.32 & 0.23 & 0.02 & 2& 0.68 \\
90842   & $-$0.29 & $-$0.05 & 0.00 & 1 & $+$0.05 & 0.13 & 3 & 0.23 & $+$0.24 &  0.16 & 2 & 0.22 & 0.09 & 9 & 0.26 & 0.44 & 0.18 & 2& 0.63 \\
146316  & $-$0.30 & $+$0.01 & 0.00 & 1 & $+$0.10 & 0.18 & 3 & 0.30 & $+$0.28 &  0.28 & 2 & 0.35 & 0.12 & 6 & 0.34 & 0.46 & 0.12 & 2& 9999 \\
15613   & $-$0.04 & $+$0.17 & 0.00 & 1 & $+$0.05 & 0.18 & 3 & 0.20 & $+$0.16 &  0.06 & 2 & 0.22 & 0.20 & 9 & 0.29 & 0.32 & 0.18 & 2& 0.55 \\
69465   & $-$0.42 & $+$0.21 & 0.00 & 1 & $+$0.01 & 0.21 & 3 & 0.19 & $+$0.27 &  0.07 & 2 & 0.15 & 0.14 & 9 & 0.15 & 0.35 & 0.14 & 2& 0.51 \\
86281   & $-$0.21 & $+$0.07 & 0.00 & 1 & $+$0.06 & 0.33 & 3 & 0.21 & $+$0.17 &  0.01 & 2 & 0.20 & 0.16 & 9 & 0.28 & 0.41 & 0.16 & 2& 0.55 \\
91298   & $-$0.20 & $+$0.07 & 0.00 & 1 & $+$0.24 & 0.11 & 3 & 0.30 & $+$0.37 &  0.03 & 2 & 0.39 & 0.10 & 7 & 0.42 & 0.53 & 0.22 & 2& 0.60 \\
9946    & $+$0.00 & $+$0.29 & 0.00 & 1 & $+$0.16 & 0.14 & 3 & 0.17 & $+$0.32 &  0.12 & 2 & 0.28 & 0.11 & 7 & 0.30 & 0.48 &   -- & 1& 0.62 \\
100719  & $-$0.70 & $-$0.05 & 0.00 & 1 & $+$0.01 & 0.16 & 3 & 0.18 & $+$0.14 &  0.07 & 2 & 0.23 & 0.13 & 9 & 0.19 & 0.25 & 0.08 & 2& 0.59 \\\hline
$<$[X/Fe]$>_{\rm 1P}$&$-$0.25&0.11&&    & 0.05    &      &   & 0.23 &  0.24   &       &   & 0.25 &      &   & 0.27 & 0.39 &      &  & 0.57 \\
$\pm$               &0.05   &0.03&&    & 0.02    &      &   & 0.01 &  0.02   &       &   & 0.02 &      &   & 0.01 & 0.02 &      &  & 0.02 \\
$\sigma$            &0.21   &0.13&&    & 0.07    &      &   & 0.05 &  0.10   &       &   & 0.07 &      &   & 0.06 & 0.10 &      &  & 0.07 \\
$<$[X/Fe]$>_{\rm 2P}$&$-$0.10&0.14&&    & 0.19    &      &   & 0.29 &  0.26   &       &   & 0.29 &      &   & 0.30 & 0.41 &      &  & 0.53 \\
$\pm$               &0.06   &0.06&&    & 0.04    &      &   & 0.03 &  0.03   &       &   & 0.04 &      &   & 0.04 & 0.05 &      &  & 0.03 \\
$\sigma$            &0.15   &0.14&&    & 0.11    &      &   & 0.07 &  0.08   &       &   & 0.10 &      &   & 0.10 & 0.12 &      &  & 0.07 \\\hline
\enddata
\end{deluxetable*}
%%%%%%%%%%%%%%%%%%%%%%%%%%%%%%%%%%%%%%%%%%%%%%%%%%%%%%%%%%%%%%%%%%%%%%%%%%%%%%%%%%%%%%%%%%%%%%%%%%%% 

The inferred chemical abundances are listed in
Tables~\ref{tab:NaToSc}-\ref{tab:TiToNi}-\ref{tab:CuToEu}. 
Internal uncertainties to these abundances introduced by errors in the
atmospheric parameters were estimated by varying the stellar parameters, one at a
time, by \teff/\logg/[A/H]/\vmicro=$\pm$50\,K/$\pm$0.20\,cgs/$\pm$0.10 dex/$\pm$0.20\,\kmsec.
The contribution due to the limits of our spectra, e.g.\ the finite S/N that
affects the measurements of EWs and the spectral synthesis
($\sigma_{\rm S/N}$), was estimated as the average rms listed 
in Tables~\ref{tab:NaToSc}-\ref{tab:TiToNi}-\ref{tab:CuToEu} divided
by the square root of the typical number of the analysed spectral
features. For those elements with only one line available, a larger
error of 0.15~dex has been adopted. 
Our error budget is listed in Table~\ref{tab:err}. The total
uncertainty $\sigma_{\rm total}$ is the quadratic 
sum of the errors introduced by the individual contributions.

%%%%%%%%%%%%%%%%%%%%%%%%%%%%%%%%%%%%%%%%%%%%%%%%%%%%%%%%%%%%%%%%%%%%%%%%%%%%%%%%%%%%%%%%%%%%%%%%%%%% 

\begin{table*}
\centering
\caption{Sensitivity of the derived abundances to the uncertainties in
  atmospheric parameters and uncertainties due to the limited S/N of
  our spectra. We reported the total internal uncertainty
  ($\sigma_{\rm total}$) associated to individual stars obtained by
  the quadratic sum of all the contributors to the
  error.}\label{tab:err} 
\begin{tabular}{lccccccc}\hline\hline
&\colhead{$\Delta$\teff} &\colhead{$\Delta$\logg}&\colhead{$\Delta$\vmicro}&\colhead{$\Delta$[A/H]} & \colhead{$\Delta$\teff}&\colhead{$\sigma_{\rm S/N}$}&\colhead{$\sigma_{\rm total}$}\\
      &\colhead{$\pm$100~K}    & \colhead{$\pm$0.20}   &\colhead{$\pm$0.20~\kmsec}& \colhead{$\pm$0.10~dex}& \colhead{$\pm$50~K}    &                              &                   \\\hline
$\rm {[Na/Fe]}$          & $\pm$0.01   & $\mp$0.03  & $\pm$0.05  & $\mp$0.01 & $\pm$0.01  &$\pm$0.05         & $\pm$0.08 \\
$\rm {[Mg/Fe]}$          & $\mp$0.01   & $\mp$0.03  & $\pm$0.01  & $\mp$0.01 & $\mp$0.00  &$\pm$0.15         & $\pm$0.15   \\
$\rm {[Al/Fe]}$          & $\pm$0.06   & $\mp$0.00  & $\mp$0.01  & $\mp$0.01 & $\pm$0.03  &$\pm$0.04         & $\pm$0.05 \\
$\rm {[Si/Fe]}$          & $\mp$0.10   & $\pm$0.03  & $\pm$0.07  & $\pm$0.01 & $\mp$0.05  &$\pm$0.03         & $\pm$0.10    \\
$\rm {[Ca/Fe]}$          & $\pm$0.03   & $\mp$0.04  & $\mp$0.00  & $\mp$0.01 & $\pm$0.02  &$\pm$0.04         & $\pm$0.06 \\
$\rm {[Sc/Fe]}$\,{\sc ii}& $\pm$0.07   & $\mp$0.01  & $\mp$0.00  & $\mp$0.01 & $\pm$0.04  &$\pm$0.06         & $\pm$0.07    \\
$\rm {[Ti/Fe]}$\,{\sc i} & $\pm$0.08   & $\mp$0.02  & $\pm$0.01  & $\mp$0.02 & $\pm$0.04  &$\pm$0.03         & $\pm$0.06    \\
$\rm {[Ti/Fe]}$\,{\sc ii}& $\pm$0.05   & $\mp$0.02  & $\mp$0.01  & $\mp$0.01 & $\pm$0.03  &$\pm$0.04         & $\pm$0.06    \\
$\rm {[V/Fe]}$           & $\pm$0.09   & $\mp$0.01  & $\pm$0.00  & $\mp$0.02 & $\pm$0.05  &$\pm$0.06         & $\pm$0.08    \\
$\rm {[Cr/Fe]}$\,{\sc i} & $\pm$0.06   & $\mp$0.02  & $\mp$0.01  & $\mp$0.01 & $\pm$0.03  &$\pm$0.06         & $\pm$0.07    \\
$\rm {[Cr/Fe]}$\,{\sc ii}& $\pm$0.02   & $\mp$0.01  & $\pm$0.02  & $\mp$0.01 & $\pm$0.01  &$\pm$0.06         & $\pm$0.07    \\
$\rm {[Mn/Fe]}$          & $\pm$0.10   & $\pm$0.02  & $\mp$0.02  & $\mp$0.01 & $\pm$0.05  &$\pm$0.02         & $\pm$0.06  \\
$\rm {[Fe/H]}$\,{\sc i}  & $\pm$0.07   & $\pm$0.01  & $\mp$0.09  & $\pm$0.01 & $\pm$0.03  &$\pm$0.01         & $\pm$0.10   \\
$\rm {[Fe/H]}$\,{\sc ii} & $\mp$0.09   & $\pm$0.10  & $\mp$0.05  & $\pm$0.03 & $\mp$0.05  &$\pm$0.03         & $\pm$0.13   \\
$\rm {[Co/Fe]}$          & $\mp$0.00   & $\pm$0.02  & $\pm$0.05  & $\pm$0.00 & $\pm$0.00  &$\pm$0.08         & $\pm$0.10    \\
$\rm {[Ni/Fe]}$          & $\mp$0.03   & $\pm$0.03  & $\pm$0.03  & $\pm$0.01 & $\mp$0.01  &$\pm$0.03         & $\pm$0.05    \\
$\rm {[Cu/Fe]}$          & $\pm$0.11   & $\mp$0.00  & $\mp$0.09  & $\mp$0.03 & $\pm$0.07  &$\pm$0.15         & $\pm$0.19    \\
$\rm {[Zn/Fe]}$          & $\mp$0.12   & $\pm$0.04  & $\mp$0.02  & $\pm$0.02 & $\mp$0.06  &$\pm$0.15         & $\pm$0.17    \\
$\rm {[Y/Fe]}$\,{\sc ii} & $\pm$0.08   & $\mp$0.02  & $\mp$0.04  & $\mp$0.01 & $\pm$0.04  &$\pm$0.14         & $\pm$0.15    \\
$\rm {[Zr/Fe]}$\,{\sc ii}& $\pm$0.03   & $\pm$0.07  & $\mp$0.02  & $\pm$0.02 & $\pm$0.02  &$\pm$0.15         & $\pm$0.17    \\
$\rm {[Ba/Fe]}$\,{\sc ii}& $\pm$0.01   & $\pm$0.06  & $\mp$0.16  & $\pm$0.03 & $\pm$0.00  &$\pm$0.07         & $\pm$0.19    \\
$\rm {[La/Fe]}$\,{\sc ii}& $\pm$0.03   & $\pm$0.07  & $\mp$0.02  & $\pm$0.01 & $\pm$0.01  &$\pm$0.06         & $\pm$0.10    \\
$\rm {[Pr/Fe]}$\,{\sc ii}& $\pm$0.00   & $\pm$0.08  & $\mp$0.01  & $\pm$0.03 & $\mp$0.00  &$\pm$0.15         & $\pm$0.17    \\
$\rm {[Nd/Fe]}$\,{\sc ii}& $\pm$0.12   & $\mp$0.02  & $\mp$0.02  & $\mp$0.00 & $\pm$0.06  &$\pm$0.14         & $\pm$0.15    \\
$\rm {[Eu/Fe]}$\,{\sc ii}& $\mp$0.01   & $\pm$0.08  & $\pm$0.00  & $\pm$0.04 & $\mp$0.01  &$\pm$0.15         & $\pm$0.17   \\
\hline
\end{tabular}
\end{table*}
%%%%%%%%%%%%%%%%%%%%%%%%%%%%%%%%%%%%%%%%%%%%%%%%%%%%%%%%%%%%%%%%%%%%%%%%%%%%%%%%%%%%%%%%%%%%%%%%%%%% 

%%_________________________________________________________________________
%%
   \begin{figure*}
   \centering
   \includegraphics[width=0.998\textwidth]{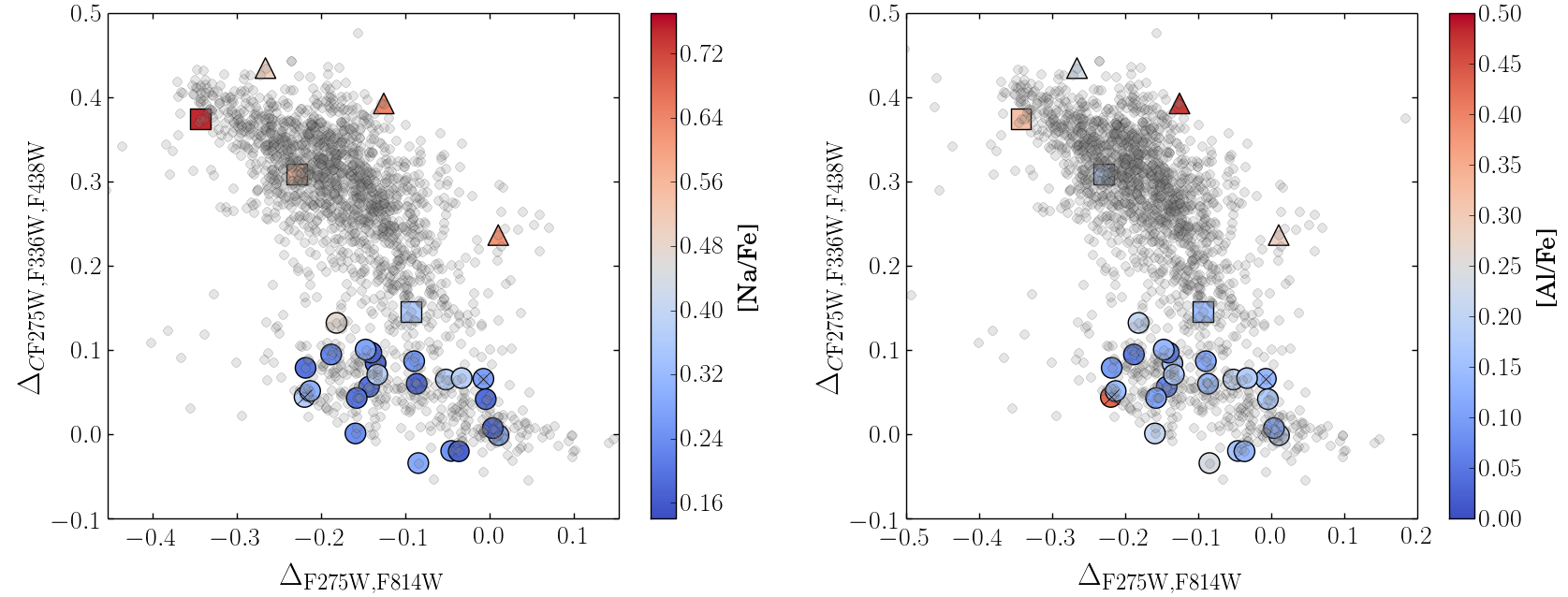}
      \caption{Location of our spectroscopic targets along the ChM of
        47~Tucanae from \citet{Mil17}. The color of each star is
        indicative of the inferred chemical abundances, as illustrated
        in the color bar. In these two plots we illustrate the
        chemical pattern of Na and Al abundances relative to Fe.} 
        \label{fig:map_light}
   \end{figure*}
%%__________________________________________________________________________
%%

%%_________________________________________________________________________
%%
   \begin{figure}
   \centering
   \includegraphics[width=0.475\textwidth]{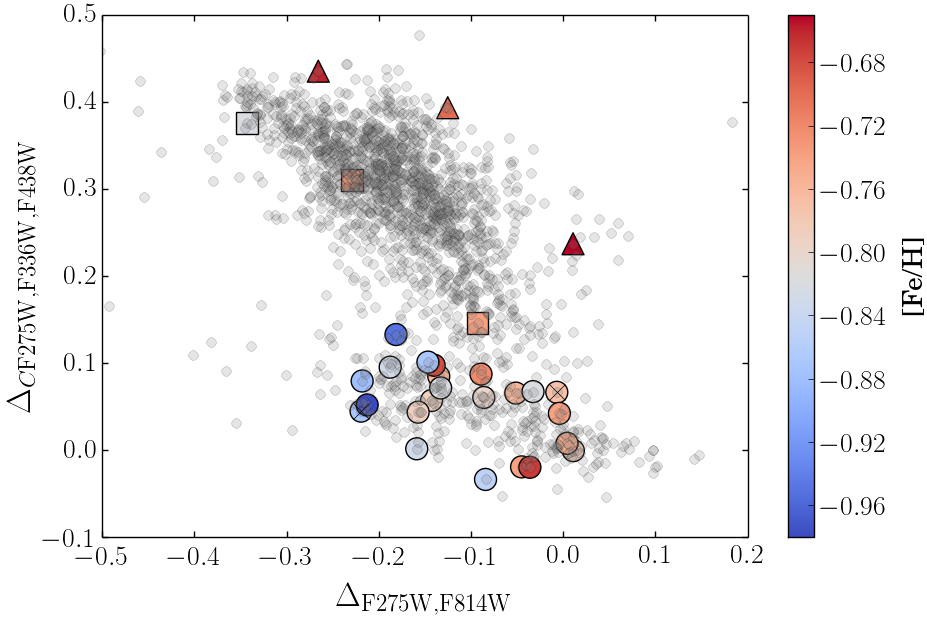}
      \caption{As Figure~\ref{fig:map_light}, but for [Fe/H]. } 
        \label{fig:map_fe}
   \end{figure}
%%__________________________________________________________________________
%%

%%_________________________________________________________________________
%%
   \begin{figure}
   \centering
   \includegraphics[width=0.48\textwidth]{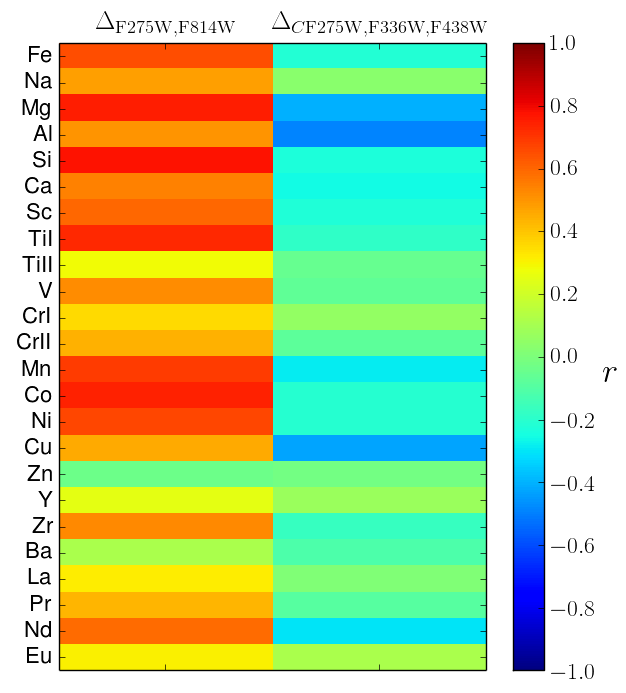}
      \caption{Chemical abundances ${\rm log} \epsilon$(X) as a
        function of \x\ and \y\ values on the ChM of 47~Tucanae. The
        color code is indicative of the Spearman correlation 
        coefficient ($r$). The coefficients and their associated significance are
        listed in Table~\ref{tab:spearman}.
      }  
        \label{fig:corr}
   \end{figure}
%%__________________________________________________________________________
%%

%%_________________________________________________________________________
%%
   \begin{figure}
   \centering
   \includegraphics[width=0.48\textwidth]{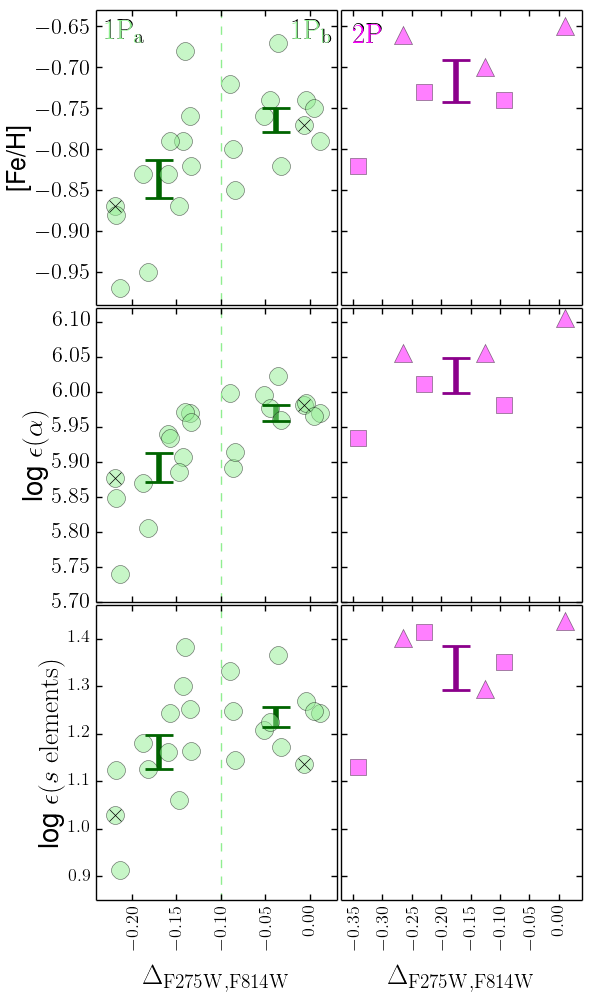}
      \caption{From top to bottom, [Fe/H] and absolute chemical abundances
        for all the $\alpha$ and the $s$ elements 
        as a function of \x\ values on the ChM of 47~Tucanae. Left and
        right panels represent 1P and 2P stars, respectively. The
        dashed line on the left panel at \x=$-$0.10 separates the two
        groups of $\rm {1P_{a}}$ and $\rm {1P_{b}}$ stars (see text
        for details). The plotted error bars represent the dispersion
        of the values in each group divided by the square root of the
        number of stars. Symbols are as in Figure~\ref{fig:chm}.} 
        \label{fig:fig4}
   \end{figure}
%%__________________________________________________________________________
%%

%%%%%%%%%%%%%%%%%%%%%%%%%%%%%%%%%%%%%%%%%%%%%%%%%%%%%%%%%%%%%%%%%%%%%%%%%%%%%%%%%%%%%%%%%%%%%%%%%%%% 
\begin{table}
\centering
\caption{Spearman correlation coefficient ($r$), and associated
  significance, between chemical abundances log$\epsilon$(X) and the
  ChM \x\ values. Lower \% reflects a more important
  correlation.}\label{tab:spearman} 
\begin{tabular}{lcc}
\hline\hline
Element         &\colhead{$r$} &\colhead{Significance (\%)} \\\hline  
Fe           &  0.62     &  1.0  \\
Na           &  0.47     &  1.7  \\
Mg           &  0.77     &  0.8  \\
Al           &  0.48     &  5.0 \\
Si           &  0.73     &  0.0  \\
Ca           &  0.58     &  4.8  \\
Sc\,{\sc ii} &  0.60     &  0.2  \\
Ti\,{\sc i}  &  0.66     &  3.4  \\
Ti\,{\sc ii} &  0.44     &  6.2  \\
V            &  0.55     &  1.2  \\
Cr\,{\sc i}  &  0.43     &  4.7  \\
Cr\,{\sc ii} &  0.49     &  0.3  \\
Mn           &  0.71     &  0.1  \\
Co           &  0.73     &  0.0  \\
Ni           &  0.62     &  0.1  \\
Cu           &  0.50     &  1.5  \\
Zn           &  0.17     & 26.9  \\
Y\,{\sc ii}  &  0.29     & 19.0  \\
Zr\,{\sc ii} &  0.52     & 17.2  \\
Ba\,{\sc ii} &  0.15     & 35.9  \\
La\,{\sc ii} &  0.31     & 20.0  \\
Pr\,{\sc ii} &  0.46     & 17.1  \\
Nd\,{\sc ii} &  0.60     &  0.2  \\
Eu\,{\sc ii} &  0.62     & 12.2  \\ \hline
\end{tabular}
\end{table}

%%%%%%%%%%%%%%%%%%%%%%%%%%%%%%%%%%%%%%%%%%%%%%%%%%%%%%%%%%%%%%%%%%%%%%%%%%%%%%%%%%%%%%%%%%%%%%%%%%%% 

\section{Chemical abundances along the ChM  of 47~Tucanae}\label{sec:results}

Overall, for our 29 stars we obtain a mean iron abundance of
[Fe/H]=$-$0.78$\pm$0.02~dex (rms=0.08~dex), consistent with the value
of [Fe/H]=$-$0.72~dex listed in \citet{Harris10}. As typical of
Population~II stars, 47~Tucanae stars are $\alpha$~enhanced in all the
analysed $\alpha$ elements, namely Mg, Si, Ca, and Ti.   

In this section we analyse the chemical abundance pattern along the ChM of 47~Tucanae. 
First, we discuss on the 1P/2P chemical differences, and then the
abundances internal to the 1P.

\subsection{Light elements}

As shown in Figure~\ref{fig:chm}, the ChM of 47~Tucanae clearly
displays a sharp separation between 1P and 2P stars occurring at
\y$\sim$0.1. Although our sample was not specifically designed to
analyse 2P stars, it includes six giants belonging to this stellar population. 
From the average chemical abundances
listed in Tables~\ref{tab:NaToSc}-\ref{tab:TiToNi}-\ref{tab:CuToEu},
 larger differences are observed in the elements involved in the hot-H
 burning processes, namely Na and Al, as expected for 2P stars in
 GCs. The six 2P stars are significantly enhanced in  
 sodium by $\Delta$[Na/Fe]$_{\rm 2P - 1P}$=0.31$\pm$0.06 and mildly enhanced in aluminum by
 $\Delta$[Al/Fe]$_{\rm 2P - 1P}$=0.11$\pm$0.05. 
 
No obvious variations are observed in neither [Mg/Fe] or [Si/Fe],
being the observed dispersions similar to those expected from
observational errors (see Table~\ref{tab:err}). Small rms values in
these chemical species have been also reported in
\citet{carretta09b}. 

For both Na and Al, the observed spread in each population is larger
for the 2P stars, consistent with the large range in \y\ of these
stars \citep[see][]{Mar19a}. Looking at the distribution of Na and Al
along the ChM, as displayed in Figure~\ref{fig:map_light}, the most 
Na rich star (\#92267) is located on the ChM region associated with
the most  Na (and He)-enriched (O-depleted) stars. Our 2P star with the
lowest \y\ value (\#41968) is also the less abundant in Na among the
analysed 2P stars. 

The observed variations in Al are smaller than those in Na (right
panel of Figure~\ref{fig:map_light}). We notice that one star located
on the {\it redder-anomalous} sequence on the ChM, namely \#116624, has
the highest Al abundance in our sample ([Al/Fe]=0.47~dex). 
Star \#8672, which we treat as a binary, has also a relatively high Al.

Among 1P stars we do not find significant variations in the [Na/Fe]
and [Al/Fe] abundance ratios. This is similar to what was previously
found among 1P stars in NGC~3201, and rules out the possibility that the
range in \x\ of the 1P stars is due to internal variations in He
\citep[see][]{Mar19b}. 
We notice here that the {\it ground level} of [Na/Fe] abundances in
47~Tucanae is relatively high, being $<$[Na/Fe]$>_{\rm
  1P}$=0.26$\pm$0.02~dex \citep[with similar values also found in
previous analysis, e.g.,][]{carretta09a}. 

\subsection{Chemical abundances along the 1P population}

In this section, we consider stars with \y\ values lower than $+$0.133
mag, a sample that includes 23 1P stars. The average Fe abundance
ratios for these stars is [Fe/H]$_{\rm 1P}$=$-$0.80$\pm$0.02~dex
(rms=0.08). The distribution of 1P stars along the {\it x} axis of the
ChM is consistent with two main over-densities, separated at \x$\sim
-$0.1. By dividing the sample into two groups we obtain [Fe/H]$_{\rm
  1P}$=$-$0.84$\pm$0.02~dex (rms=0.08, 12 stars), and [Fe/H]$_{\rm
  1P}$=$-$0.76$\pm$0.02~dex (rms=0.05, 11 stars), for stars with \x\
lower (hereafter 1P$_{\rm a}$) and higher (hereafter 1P$_{\rm b}$)
than $-$0.1, respectively. Hence, the mean difference we get is
$\Delta$[Fe/H]=0.08$\pm$0.03~dex, an almost 3~$\sigma$ difference. 

From Figure~\ref{fig:map_fe}, we note that the distribution of iron
abundances looks less homogeneous in the stars with lower \x\ values,
questioning the possible presence of two separated groups rather than
a continuous distribution. 
Indeed, $\rm {1P_a}$ stars have a significantly larger dispersion in
Fe than $\rm {1P_b}$ ones, suggesting that this group could be more
heterogeneous. If we consider only stars with \x$< -$0.17 we obtain
[Fe/H]$_{\rm 1P}$=$-$0.90$\pm$0.03~dex (rms=0.06, 5 stars), and the
difference in Fe of 0.14~dex with the 1P$_{\rm b}$ stars, which is at
a $>3~\sigma$ level. 

The observed differences in iron are smaller than those typically
inferred for the blue- and red-RGB stars in most Type~II GCs
\citep[see][]{Mar19a}, and may be more difficult to detect. Our
relatively high-S/N and high-resolution UVES spectra 
allow us to detect such difference.

Within our uncertainties, chemical abundances relative to Fe ([X/Fe])
for all elements are generally consistent with homogeneous content.  
By comparing the observed rms associated to the mean average
abundances, as listed in
Tables~\ref{tab:NaToSc}-\ref{tab:TiToNi}-\ref{tab:CuToEu}, with the
estimated errors, it appears that in most cases our expected
uncertainties are higher. This suggests that our estimated errors
might be overestimated, in particular for the $n$-capture elements.  

Figure~\ref{fig:corr} shows a representation of the Spearman
correlation for all the analysed log$\epsilon$(X) abundances as a
function of the ChM \x\ and \y\ values, for the 1P stars. The
correlation coefficients $r$ are listed in Table~\ref{tab:spearman}. 
With each $r$ value we have associated a significance, which we
obtained from a Monte Carlo simulation of 1000 realizations of our
dataset composed of 23 stars. In each realization we have assumed the
observed \x\ and a uniform Fe abundance with the associated error
estimates of Table~\ref{tab:err}, and derived the slope. We have
calculated the fraction of realizations where the slope is higher than
the observed one, and assumed this value as the probability that the
slope is due to randomness. 

From the significance values \% listed in Table~\ref{tab:spearman}, we
note that not all correlations are significant, in particular larger 
uncertainties are associated with the {\it neutron}-capture elements.
However, all the absolute abundances appear to be positively
correlated with the \x\ values. The negative correlation with \y\ is a
reflection of the mild negative correlation between \x\ and \y\ values
for 1P stars.  
The general correlation between each log$\epsilon$(X) abundance and
the \x\ values supports the presence of a small spread in the overall
metallicity among the 1P stars. 

In Figure~\ref{fig:spectra} we show a portion spectrum from two 1P
stars with similar atmospheric parameters, and Fe abundances of
[Fe/H]=$-$0.76 and [Fe/H]=$-$0.88~dex. Overall, the spectral features
of the star at higher \x, namely \#11746, are more consistent with
higher metallicity than the spectrum of star \#112499. 

As discussed in \citet{Mar19b} for NGC~3201, also in 47~Tucanae we
exclude a variation in the He content intrinsic to the 1P stars as
responsible for the \x\ spread, which should be as high as
0.054$\pm$0.006s in mass fraction \citep{Mil18}, without any
corresponding enhancement in other chemical species, such as Na or
Al. 
Another effect to be considered is that a change in Y will affect the
metal-to-hydrogen ratio, Z/X, since X + Y + Z = 1. Hence, at constant
Z, a He-rich star will appear to be slightly more metal-rich than a
He-normal star. Noting that a {\it pure} He enhancement shifts stars
towards lower \x\ values along the ChM \citep{Mil15, Mar19a}, the
possibility that increased metallicities are the result of He
enhancements can be ruled out, as the stars that should be enhanced in
He (at lower \x), are metal-poorer. 

\citet{Mar19b} also discussed the possibility that the presence of
binaries among 1P stars can translate in spurious lower metallicity
abundances for stars with lower \x.  
Most of our stars for 47~Tucanae have been observed spanning over a
relatively large range in time to spot for possible variations in the
RVs values indicative of binarity (see last column of
Table~\ref{tab:targets}).  
As already mentioned in Section~\ref{sec:spec_data}, only two stars
have significantly larger dispersions in their RVs, but we cannot
exclude the presence of close binaries in our sample with lower
(undetectable) variations in the RVs. 
However, we enphasize here that \citet{Mar19b} pointed out that
significant effects in the emerging spectra are present for
giant-giant pairs, which is unlikely to exist in such a large fraction
to account for the bluer 1P stars. 

We conclude that the 1P of 47~Tucanae is not chemically homogeneous. A
small intrinsic variation in the overall metallicity is present among
1P stars of this globular cluster.

\subsection{Chemical abundances in second population and anomalous stars}

In this section we investigate the overall chemical pattern, besides
the light elements, between the 1P and 2P stars, which include also
three stars associated with the {\it anomalous} component. 

The first thing to note is that, by looking at the ChM in
Figure~\ref{fig:map_fe}, the {\it anomalous} stars have somewhat higher Fe
abundance, with the most Fe rich stars being the {\it anomalous} stars
\#22386 and \#73063, with [Fe/H]=$-$0.65~dex and [Fe/H]=$-$0.66~dex,
respectively.
Comparing the average Fe abundance for the three {\it anomalous}
stars, namely [Fe/H]=$-$0.67$\pm$0.02~dex (rms=0.03~dex), with that of
the stars belonging to the blue {\it normal} component (26 stars),
with average abundance of [Fe/H]=$-$0.80$\pm$0.01~dex (rms=0.07~dex),  
it results that the {\it anomalous} stars in 47~Tucanae are enhanced
in iron. This is similar to what is found in most Type~II GCs. 

However, the presence of internal variations in Fe among 1P stars
imposes a more careful analysis of possible Fe enhancements in {\it
  anomalous} stars.  
By plotting the abundances as a function of the \x\ values of the ChM,
we obtain the chemical pattern displayed in Figure~\ref{fig:fig4}. 
In this figure we show the average [Fe/H] abundances, as well as the
average absolute abundances from all the $\alpha$ and the $s$
elements, weighted for the corresponding observational errors. 

Compared with the 1P, we first notice some hints of chemical enrichment in 2P stars. 
Translating into numbers, the mean chemical abundance differences
between 2P stars (all, {\it normal}, and {\it anomalous}) and 1P stars
($\rm {1P_{a}}$ and $\rm {1P_{b}}$) are listed in
Table~\ref{tab:differences}. From these numbers the chemical
enrichment of {\it anomalous} stars is clearly confirmed for all the
considered chemical species above 3~$\sigma$ level, except for the
$s$~elements, which have, however, larger uncertainties. 

When dividing 1P stars into $\rm {1P_{a}}$ and $\rm {1P_{b}}$, the
abundances of 2P {\it normal} stars are closer to the abundances of
the $\rm {1P_{b}}$ stars 
than to those of the $\rm {1P_{a}}$ ones. This pattern is best
illustrated by considering the $\alpha$ elements, that have lower
associated uncertainties. Similar trends are observed for all the
other considered chemical species. 

Finally, we note that the dispersion in chemical abundances for 2P
stars, except for the elements involved in the hot-H burning, are
smaller than those associated with the 1P ones. As an example, the
[Fe/H] abundance dispersion is 0.08~dex for the 26 1P and 0.06~dex for
the six 2P stars. The difference is larger if we divide the 2P
population in {\it normal} and {\it anomalous}, with dispersions of
0.05 and 0.03~dex, respectively. 

%%%%%%%%%%%%%%%%%%%%%%%%%%%%%%%%%%%%%%%%%%%%%%%%%%%%%%%%%%%%%%%%%%%%%%%%%%%%%%%
    
\begin{table*}
\caption{Differences in [Fe/H] and in the absolute abundances log$\epsilon$(X) for all the
  analysed elements, the $\alpha$ and $s$ elements between the
  stellar populations of 47~Tucanae. We have listed the differences
  between the 2P stars (all, {\it normal},  and {\it anomalous}) and
  1P stars subdivided into \x$< -0.10$ ($\rm {1P_{a}}$) and $>-0.10$
  ($\rm {1P_{b}}$).}\label{tab:differences} 
\begin{tabular}{c c ccc}
\hline\hline
                              &[Fe/H] &  \multicolumn{3}{c}{$\Delta$log$\epsilon$(X)}             \\
                              &       &all elements  & $\alpha$ elements & $s$ elements   \\\hline
2P (all) $-$ $\rm {1P_{a}}$              &0.12$\pm$0.04 &0.13$\pm$0.05 & 0.13$\pm$0.03 & 0.18$\pm$0.06     \\
2P (all) $-$ $\rm {1P_{b}}$              &0.05$\pm$0.03 &0.07$\pm$0.04 & 0.05$\pm$0.03 & 0.10$\pm$0.05     \\
2P ({\it normal}) $-$ $\rm {1P_{b}}$     &0.00$\pm$0.04 &0.02$\pm$0.04 & 0.01$\pm$0.03 & 0.06$\pm$0.09     \\
2P ({\it anomalous}) $-$ $\rm {1P_{b}}$  &0.09$\pm$0.02 &0.11$\pm$0.02 & 0.10$\pm$0.02 & 0.14$\pm$0.05     \\ \hline
\end{tabular}
\end{table*}

%%%%%%%%%%%%%%%%%%%%%%%%%%%%%%%%%%%%%%%%%%%%%%%%%%%%%%%%%%%%%%%%%%%%%%%%%%%%%%%

\section{Discussion}\label{sec:discussion}

In this section we explore the chemical abundance pattern of
47~Tucanae in the context of a plausible star formation history for
its stellar populations. For simplicity, this discussion follows the
different populations that have been detected through a list of
subsections. 

\subsection{1P stars}

1P stars themselves are not homogeneous in metallicity reflecting in a
spread along the the {\it x} axis of the ChM. 
We do not find any ``indirect'' evidence for the presence of He variations within these stars.
Indeed, the [Na/Fe] and [Al/Fe] abundances are consistent with uniform
composition, with no evidence of chemical enrichment in none of these
elements involved in the hot-H burning. 
In addition, we can exclude the variation in metals to be a
consequence of a change in helium (Y) and, consequently to Z/X, as in
this case we would expect the opposite trend with the \x\ values. 

As shown for another GC, namely NGC\,3201, non-interacting binaries
with mass ratio {\it q}$\gtrsim$0.8 can produce a sizable shift
towards low \x\ values \citep{Mar19b}.  
Although the presence of other binaries in our sample cannot be
excluded, only two 1P stars can be classified as binaries from RVs,
with one of them displaying a low \x\ value. 
Simulated spectra of giant–giant pairs are consistent with higher
temperature and/or lower metallicity (see the discussion in
\citet{Mar19b} for further details). In such a case the stars with
lower metallicity and low \x\ values would be binaries or blue
stragglers.  
Since no evidence exists supporting large fractions of such binaries
to account for the entire color spread of 1P stars in GCs
\citep{APM12bin}, our analysis for 47~Tucanae corroborates the idea of
a {\it genuine} spread in the overall metallicity among these stars. 

As photometrically shown by \citet{Legnardi22}, a spread in metals of
$0.087\pm0.009$~dex among the 1P stars is sufficient to generate a
$\sim 0.1$~mag spread in the $\rm {m_{F275W}-m_{F814W}}$ color of the
RGB stars of 47~Tucanae.  
In this context, we note that the ChM of M-dwarfs constructed with
F606W filter of the UVIS/WFC3 camera on board {\it HST} and the F115W
and F322W2 IR filters from the James Webb Space Telescope also
displays an elongated 1P distribution, consistent with an internal
variation in metals of $\sim$0.10~dex  \citep{Mil23}. 

The evidence of metallicity variations among both the RGB 1P stars and
the unevolved low-mass MS stars, would suggest that the variation in
metals reflects the chemical composition of the cloud from which 1P
stars formed. 

\subsection{2P {\it normal} stars}

As expected, the 2P stars are enhanced in Na, at different degrees
depending on their \y\ value. The {\it normal} 2P star with highest
\y\ which also displays the lowest \x\ value in our sample, is the
most Na rich.  

Although we warn about the small number of analysed stars, it is
interesting to note that chemical abundances of the 2P stars are
overall more similar to that of the 1P stars with higher \x\ values,
namely what we have defined as the $\rm {1P_{b}}$ population. By
selecting only the 2P stars belonging to the {\it normal} component,
the chemical abundances (with the exception of Na and Al) are
consistent with those of the $\rm {1P_{b}}$ stars. 

Furthermore, the 2P stars do not seem to show the same degree of
inhomogeneity as the 1P ones. This is more clear if we subdivide 2P
stars in {\it normal} and {\it anomalous}. However, we note that the
available 2P targets are, at the moment, a quite limited sample to
draw strong conclusions on this issue.

\subsection{Anomalous stars}

The three candidate {\it anomalous} stars are enhanced in the overall
metallicity. Their [Fe/H] is higher by $\sim$0.10~dex than that of the
mean [Fe/H] abundance of $\rm {1P_{b}}$ and the {\it normal} 2P, and
by $\sim$0.20~dex than the $\rm {1P_{a}}$ stars. On the other hand, we
do not find any evidence for a significant enrichment in the $s$
elements for these stars, as observed for many Type~II GCs \citep[see
Table~10 in][]{Mar15}.  
However, we cannot exclude that, due to the larger observational
errors associated with the abundances of these elements, some
enrichment could be present, possibly lower than that observed in some
Type~II GCs, such as NGC\,1851 and NGC\,5286. 

Furthermore, our {\it anomalous} stars are all enhanced in Na, in
agreement with their relatively-high values of \y\ compared to the 1P
stars. This suggests that the {\it anomalous} population of 47~Tucanae
could have entirely formed from an ISM which was already enriched in
the H-burning products.  

It is interesting to note that our findings qualitatively agree with
the chemical pattern observed on the faint sub giant branch of
47~Tucanae by \citet{Mar16} who found higher N abundances for these
stars. Theoretical investigation has suggested that the faint sub
giant branch stars, the progeny of the {\it anomalous} red giants,
could be enhanced in He and in C+N+O \citep{diCriscienzo10}. While the
nitrogen abundances are consistent with a possible enhancement in He
of these faint sub-giants, no strong evidence has been found for C+N+O
enrichment. However, the predicted variation is of the order of
$\sim$0.10~dex, which is within the estimated observational errors
\citep{Mar16}. 

In the context of Type~II GCs, a similar pattern has been  shown, at
different degrees, for other clusters, including M\,22 and
$\omega$~Centauri. Both of them host indeed {\it anomalous} stars with
relatively-low Na that can be associated with an {\it anomalous} 1P;
however, the [Na/Fe]-{\it ground level} of {\it anomalous} 1P stars is
higher than that associated with the {\it normal} 1P \citep{Johnson10,
  Mar11a, Mar11b, Mar19a}. 
Other Type~II GCs, such as NGC\,1851, have a {\it more extreme}
pattern with a completely (or largely) missing {\it anomalous} 1P, and
most, if not all, {\it anomalous} stars are consistent with being
enhanced in the H-burning products, such as Na. On the other hand, the
Type~II GC NGC\,6934, besides a Na-rich {\it anomalous} population,
hosts a Na-poor {\it anomalous} component comparable with the {\it
  normal} 1P stars \citep{Mar21}. 

This discussion highlights a quite heterogeneous picture for the
chemical composition in light elements of {\it anomalous} stars in
Type~II GCs, suggesting that each cluster might have experienced a
different timing for the onset of its {\it anomalous} population, and
different nucleosynthetic processes contributing to that population.

\subsection{Interpreting the observations}

This work outlines how complex is the chemical inventory 
of 47~Tucanae, which however is not a one-off among Milky Way
GCs. Already among the 1P we observe internal variations in the
overall metal content. If these small variations are not associated to
binarity, they can be interpreted either as the result of  
inhomogeneity in the primordial cloud from which the GC formed, or
from intra-cluster chemical enrichment occurring in the first phase of
the proto-cluster evolution, before the medium is enriched in the
products of the H-burning. 
A similar spectroscopic pattern has been also observed in NGC\,3201 by
\citet{Mar19b} and, from photometric diagrams, in a large sample of
Milky Way GCs analysed by \citet{Legnardi22}.  

Since the idea of internal Fe enrichment starting from the
metal-poorer stars would dramatically exacerbate the {\it mass budget
  problem}, we favour the idea of an intrinsic inhomogeneity in the
primordial proto-cluster cloud, which is theoretically supported by
recent hydrodynamical simulations \citep{McKenzie21}. 

To understand how such a spread could have been established in the
first place, let us first quantify the iron spread in terms of
supernova yields. The 1P of 47~Tucanae has a mass of $\sim 2 \times
10^5\,\ {\rm M}_\odot$ and a metallicity $\sim 0.2$ solar. For a solar
iron abundance $Z_\odot^{\rm Fe}=0.00124$ \citep{asplund09}, this
gives $Z^{\rm Fe}=0.000248$. If half of 1P is 0.1~dex richer in iron,
this corresponds to $Z^{\rm Fe}=0.000312$, hence the iron-rich half of
1P has $\sim 6.4\,{\rm M}_{\odot}$ of iron more than the iron-poor
half. This is fairly large amount, taking into account that the
average iron product of each core collapse (CC) or Type Ia supernova
is $\sim 0.07\,{\rm M}_{\odot}$ and $\sim 0.7\,{\rm M}_{\odot}$,
respectively \citep[e.g.,][and references
therein]{Renzini2014}. Hence, these $\sim 6.4\,{\rm M}_{\odot}$ of
iron would correspond to the yield of $\sim 100$ CC supernovae or
$\sim 10$ Type~Ia supernovae, respectively. 
We conclude that it is plausible that such number of Type~Ia
supernovae from previous populations may have blown up inside the
molecular cloud going to form the 1P of 47~Tucanae, and that there was
not time enough for complete mixing of its iron 
yield. It is also possible that the material going to form the 1P was
collected from ISM regions having experienced a different number of CC
supernovae, when considering that the full iron content of the 1P is
$50\,{\rm M_{\odot}}$, or the product of $\sim 700$ CC supernovae.
A major intra-cluster chemical enrichment in metals could have
occurred at later times, responsible for the formation of the {\it
  anomalous} component. 

We notice here that \citet{Kirby23} has recently detected an internal
variation in $r$-process elements affecting only 1P stars in the GC
M\,92. This result further corroborates the occurrence of
inhomogeneous chemical pollution from SNe, which was internal to the
1P, before the formation of the 2P stars.  

The onset of 2P stars associated with the {\it normal} component
  may have occurred from the metal-richer 1P stars, the $\rm
{1P_{b}}$. This would be in agreement with the ChM morphology of
47~Tucanae, which shows the 2P main branch starting from
relatively-red \x\ values, at the location of  the metal-richer 1P.   

Noticeably, the dispersion of the 2P {\it normal} stars is smaller
than that observed in the 1P. Even though this result needs to be
confirmed over a larger sample of stars, a higher degree of
homogeneity in the overall metal content of these stars is remarkable.  
Understanding whether or not 2P stars display the same level of
inhomogeneity as the 1P stars is indeed crucial for GC formation
scenarios. If 2P {\it normal} stars are homogeneous in Fe, this will
strongly corroborate the idea that successive generations of stars
formed in a high-density environment, very likely the proto-clusters
central regions after a cooling flow process mixes and homogenizes the
ISM.  
On the other hand, if 2P normal stars will trace similar Fe variations
as the 1P, the different stellar populations would be hardly
reconcilable with multiple bursts of star formation. In this second
case, the observations would support the early disc accretion
scenario, rather than the multiple generations one. 
Future observational efforts will be certainly devoted to constrain metallicity spreads also in 2P stars, 
and, hopefully, theoretical studies based on hydrodynamical
simulations of GC formation in the early Universe will need to include
the new observational results on this issue.

\section{Summary}\label{sec:summary}

We have presented chemical abundances from UVES spectra of 29 red
giant branch stars in the GC 47~Tucanae.  
Our spectroscopic sample has been selected to prioritize the
observation of 1P stars that span a large range in \x\ values along
the ChM.  
Hence, our target sample consists of 23 1P stars, including two binary
candidates, and six 2P stars, including three stars distributed on the
{\it normal} component of the ChM, and other three on a possible {\it
  anomalous} one, as defined in \citet{Mil17}. 

Our analysis suggests that a variation in the overall metallicity of
$\sim$0.10~dex is responsible for the distribution of 1P stars along
the ChM. The average [Fe/H] abundance for 1P stars with \x$> -$0.10 is
$\sim$0.10~dex higher than that for 1P stars with lower values. The
absolute abundances of all the other analysed elements follow the same
variation of Fe, keeping constant the abundance ratios relative to
Fe. 

The abundances of the 2P stars, which include both {\it normal} and
{\it anomalous} stars, suggest that the bulk of 2P may have formed
from 1P material with higher metal content, and that the cluster
experienced a further enrichment in metals, which was responsible for
the formation of the {\it anomalous} population. The small number of
2P {\it normal} analyzed stars does not display a dispersion in metals
as large as that observed among 1P stars. 

The complex chemical inventory of 47~Tucanae, as emerged from this
work, not confined to the commonly-observed light element patterns,
further corroborates the intricate interweaving of stellar populations
in GCs. Clearly, these objects, once considered the simplest examples
of stellar systems, have experienced the interplay of different  
processes which produced the observed chemical diversity among GC stars.

%
%__________________________________________________________________
\acknowledgments
The authors thank the anonymous referee for useful discussion.
Based on observations collected at the European Organisation for
Astronomical Research in the Southern Hemisphere under ESO programme
105.20NB. 
This work has received funding from INAF Research GTO-Grant Normal
RSN2-1.05.12.05.10 - {\it Understanding the formation of globular
  clusters with their multiple stellar generations} (ref. Anna
F. Marino) of the "Bando INAF per il Finanziamento della Ricerca
Fondamentale 2022", from the European Union’s Horizon 2020 research
and innovation programme under the Marie Skłodowska-Curie Grant
Agreement No. 101034319 and from the European Union –
NextGenerationEU. MT acknowledges support from the ERC Consolidator
Grant funding scheme (project ASTEROCHRONOMETRY,
https://www.asterochronometry.eu, G.A. No. 772293).  
%__________________________________________________________________
%

\end{document}